\documentclass[structabstract]{aa} 
\usepackage{graphicx}
\begin{document}
\title{VLT/NACO near-infrared imaging and spectroscopy of N88A  in  the SMC \thanks{Based on observations
 obtained at the European Southern
Observatory,  El Paranal, Chile}}
\author{G. Testor\inst{1}$^{**}$,
 J.L. Lemaire\inst{1},\inst{2}\thanks{visiting astronomer at VLT Paranal} 
, M. Heydari-Malayeri\inst{1}, L. E. Kristensen\inst{3}, S. Diana\inst{2}, D. Field\inst{4}}
\offprints{gerard.testor@obspm.fr and jean-louis.lemaire@obspm.fr}
\institute{LERMA, UMR 8112 du CNRS, Observatoire de Paris, 92195 Meudon, France
\\
e-mail: gerard.testor@obspm.fr
\and
Universit\'e de Cergy-Pontoise, 95031 Cergy Cedex, France 
\and
Leiden Observatory, Leiden University, Niels Bohrweg 2, 2333 CA Leiden, The Netherlands
\and
Department of Physics and Astronomy, \AA{}rhus University, 8000
 \AA{}rhus C, Denmark.
}
   \date{Received ...; accepted ...}
                         \abstract
  {}
   {We present  near-infrared imaging and spectroscopic high spatial resolution observations 
  of the SMC region N88 containing the bright, excited, extincted and compact  H II region N88A of size 
  $\sim$1 pc.}
   {To investigate its stellar content and  reddening,   
 N88 was observed  using spectroscopy and imagery in the JHKs-  and L'-band  at a spatial 
 resolution of $\sim$ 0.1-0.3 $\arcsec$, using  the VLT UT4 equipped with the NAOS adaptive optics system.
 In order to attempt to establish if the origin of the infra-red (IR) excess is due to bright nebulosity,
 circumstellar material and/or local dust, we used  Ks vs J-K colour-magnitude (CM) and JHK
 colour-colour (CC) diagrams, as well as L' imagery.}
 {Our IR-data reveal in the N88 area an IR-excess fraction of $\geq$30 per cent of the detected stars, as
 well as an unprecedently detailed morphology of N88A. It consists  of  an embedded cluster
 of $\sim$ 3.5$\arcsec$ ($\sim$ 1 pc) in diameter, of at least thirteen
 resolved stars  superposed with an unusual bright continuum centered on a very bright star.
 The four brightest stars in this cluster  lie red-ward of H-K $\geq$ 0.45 mag, and
 could be classified as young stellar object (YSO) candidates. 
 Four other probable  YSO candidates are also detected in N88 along a south-north bow-shaped
 thin H$_2$ filament at $\sim$ 7$\arcsec$ east of the young central bright star. This star, 
 that we assume to be the main exciting source, could also be complex. 
 At 0.2$\arcsec$ east of this  star, a heavily embedded core is detected
 in the L'-band. This core with L' $\sim$ 14 mag and L'-K $\geq$ 4.5 mag 
 could be a  massive class I protostar candidate. The  2.12 $\mu$m H$_2$ image of N88A
 resembles a shell of diameter $\sim$ 3$\arcsec$ ($\sim$ 0.9 pc)  centered on the bright star. This shell
 is formed of three bright components,  of  which the brightest one  superposes the ionization 
 front. The line ratios of H$_{2}$ 2-1 S(1) and 1-0 S(0) relative to 1-0 S(1), as well as the 
 presence of high v lines,  are indicative  of photodissociation regions, rather than shocks.}
   {}

   
\authorrunning{G. Testor et al.}
\titlerunning{ NIR VLT/NACO observations of N88A in the SMC}
   \maketitle

\section{Introduction}
  The  Small Magellanic Cloud (SMC) is rich in  H II regions and young OB associations.
  Because of its known and  relatively small  distance ($\sim$ 65 kpc) (Kovacs 2000),
  its face-on  position relatively free from  foreground extinction  (McNamara $\&$ Feltz 1980), and 
  low internal extinction (Westerlund 1997), it is  well suited  to study
  both individual stars and  very compact objects, as well as global structures. It
  is an  ideal laboratory  for studying  the formation   and evolution of massive stars
  in a low metallicity environment.   
  Understanding the characteristics of massive stars and their interaction with their
  environment is a key problem in astrophysics. We have made some progress concerning
  the early stages of massive star formation in the galaxy, but the current knowledge about
  the early stages of massive star evolution in other galaxies is mediocre at best.
  The main  reason is that   the earliest stages of massive star evolution are deeply
  enshrouded, inaccessible in the optical wavelengths. Another reason is that these stars 
  are often members of very crowded regions.  Today, high-spatial   near-infra-red
 (NIR) resolution observations using NACO attached  to the VLT, are able to overcome
  these obstacles in  the SMC. Our search for the youngest massive stars in the
  Magellanic Clouds (MCs) (Heydari-Malayeri   $\&$ Testor 1982) led to the discovery 
  of a distinct and very rare  class of H II  regions that we called high-excitation
  compact H II "blobs" (hereafter HEBs) listed in  Testor (2001). 
  In contrast  to the ordinary H II regions of the MCs, which are extended  structures 
  spanning several arcminutes on the sky ($>$ 50 pc) and are powered by a large number 
  of hot stars, HEBs are very dense small regions, usually 2$\arcsec$ to 8$\arcsec$ 
  in diameter (0.8 to 3 pc), ionized  by one or a few massive stars and affected by local dust. 
   Two other HII  regions,  MA 1796 and MG 2 (less than  1pc across) heavily extincted and ionized
  by a small young cluster, have been  found by Stanghellini et al. (2003) in the SMC.
 
 In the present paper we focus on the peculiar   HEB LHA 115-N88A, hereafter labelled  N88A 
 of diameter $\sim$1 pc (Testor $\&$ Pakull 1985) in the extended H II region LHA 115-N88
 or N88 (Henize 1956) of
 diameter $\sim$ 10 pc. N88 lies in the  Shapley Wing of the SMC and contains the
 young cluster HW 81 ($\sim$ 0.6$\arcmin$)  (Hodge $\&$ Wright 1977).
 It is known that the SMC is made of four H I  layers with different velocities
 (McGee \& Newton 1982). This situation complicates the study, particularly in the region
 of the Shapley wing. However, the available H I observations provide helpful data for the
 study of this region. In particular, the N88 region lying at about 35$\arcmin$
 ($\sim$ 700 pc) west of N83/N84 is not apparently associated 
with the H I cloud of these H II regions (Heydari-Malayeri et al. 1988).
 N88A should be associated with the H I component of velocity +134 kms$^{-1}$ 
(McGee \& Newton 1982).

  N88A is the  brightest and  the most  
excited HEB in the MCs. It is also the most reddened H II region in these galaxies of
 low-metallicity (Heydari-Malayeri et al. 2007). Numerous optical detailed  studies
 have been made  on this object  (Testor $\&$ Pakull 1985, Wilcot 1994, Heydari-Malayeri
 et al. 1999 (hereafter HM99), Kurt et al. 1999, Testor et al. 2003). Nevertheless,
 many uncertainties remain  to understand the  true nature of N88A, such as its exciting 
 source, that still  remains   unidentified, as well as the nature of the reddening. 
  
 Israel $\&$ Koorneef (1988, 1991) have detected the presence  of molecular hydrogen
 in N88 through  H$_{2}$ emission which  is either shock-excited on a small
 scale of 0.46$\arcsec$ by stars embedded in the molecular cloud, or radiatively
 excited on a large scale (3$\arcsec$-60$\arcsec$). However, their low spectral (R=50)
 and spatial (7.5-10$\arcsec$ aperture) resolutions did not allow discrimination of these
 different processes. They described N88A  as a strong NIR source dominated by nebular 
 emission containing a strong hot dust component and noticed  that N88A has an unusual
 blue J-H colour.  In Testor et al. (2005), at low spatial resolution,   a pure H$_{2}$
 emission is detected in N88A as well as  along a south-north diffuse long  filament at
 $\sim$  6-8$\arcsec$ to the  east.
  In N66 (Henize 1956) a giant HII region in the SMC, Schmeja et al. (2009) have reported that
 most of  the H$_{2}$ emission peaks coincide with the bright component of the ionized gas
 and with compact embedded young clusters where candidate YSOs have been identified.
 Using SEST,  Israel et al. (2003) detected a  CO  molecular cloud of 1.5$\arcmin$ x 1.5$\arcmin$ 
 in the region, reporting spectra and maps of the $^{12}$CO lines J=1-0 and J=2-1.
 Stanimirovic et al. (2000)  found that the highest values of the  dust-to-gas mass ratio and
 dust temperature in the SMC are found  in N88A.

 IR studies  of a  similar size young  star formation region like the Trapezium 
 region in Orion ($\sim$0.75 x 0.75 pc)  (Lada et al. 2000)  and the more extended
 30 Doradus in the LMC (Maercker $\&$ Burton 2005) showed that during star formation,
 YSOs are associated with the circumstellar  material
 inducing IR-excess emission, and  also that the use of JHK CC
 diagrams are useful tools to detect this emission.  However, for young massive stars
 generally found in embedded clusters,  their  surrounding material destruction
 time scale is short, making their observation difficult (Bik et al. 2005, 2006).
 The most suitable wavelength to determine the nature of the  IR-excess is the
 L-band,  increasing  the IR-excess and reducing  the contribution of  extended 
 emission  from reflection nebulae and H II regions (Lada et al. 2000). IR-excess
 can be  useful to determine the origin of  the reddening in embedded young clusters.
 Martin-Hernandez et al. (2008) found in N88A,  from a  mid-IR high spatial
 resolution  Spitzer-IRS spectrum, a  rising dust continuum and PAH bands, typical 
 characteristics in H II regions. Using radio observations, Indebetouw et al. (2004)
 found that N88A is ionized by an O5 type star. 

 In the  present paper we present  the results  of JHK- and L'-band  high spatial resolution
 observations  of  N88A  and its surroundings, using adaptive optics at the VLT.
 Section 2 discusses the instrumentation  employed during these observations, and the data
 acquisition and reduction procedures used.  Section 3 describes the results and analysis
 of our observations, and  Section 4 gives our conclusions.
\begin{figure*}[t]
\begin{center}
\hspace{0cm} 
\includegraphics[width=14.0cm]{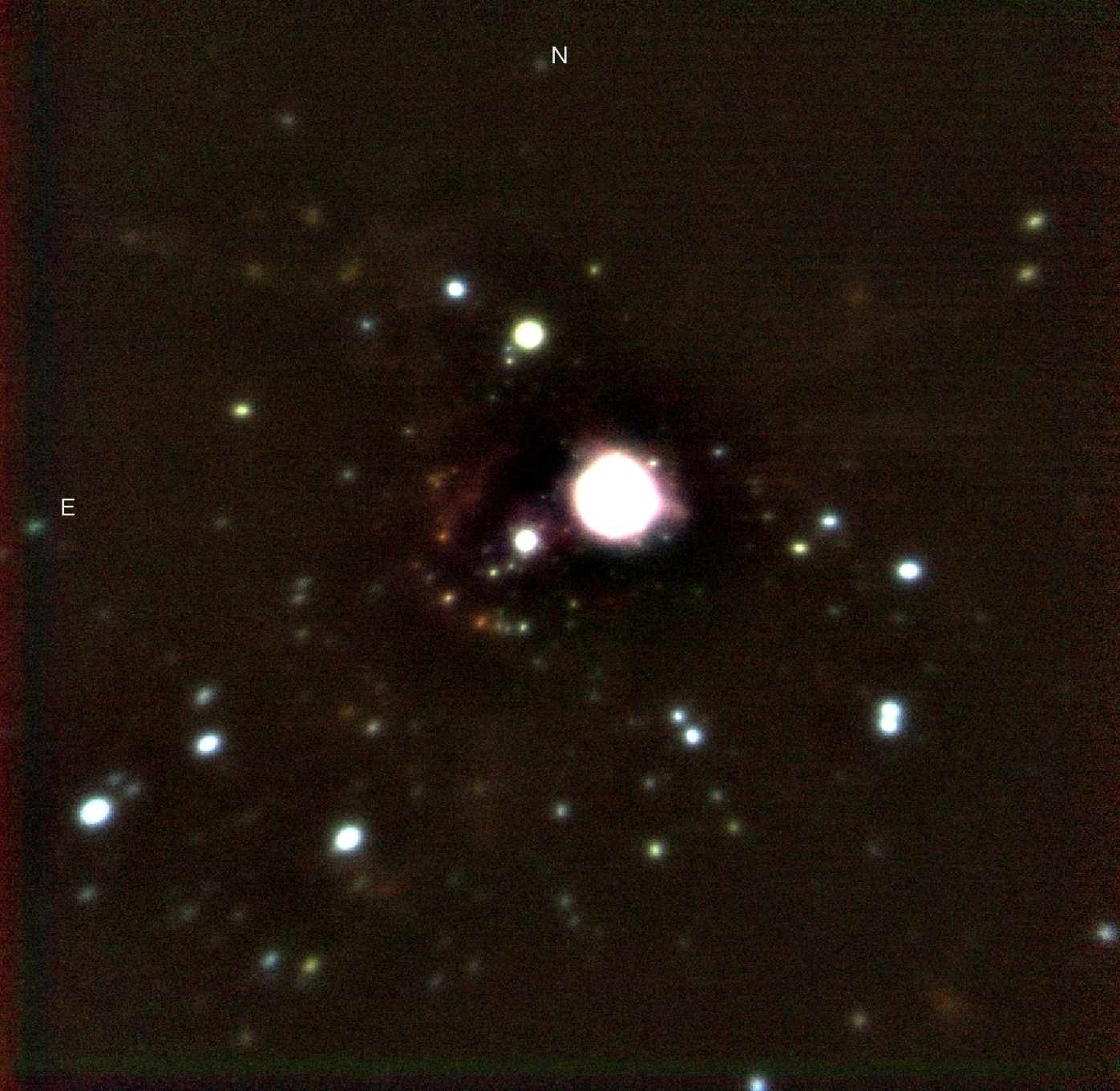}
\caption{JHK colour composite image of SMC N88 containing the bright HEB N88A.   
        The image  size extracted from the S54 camera field corresponds to 50$\arcsec$ x 50$\arcsec$ 
        (or $\sim$ 14 pc x 14 pc). The lower spatial density of faint stars 
        close to the edges of the frame delimits approximatively the cluster N88-cl. In
        the north-west quadrant the lack of stars is explained by the presence  of the molecular cloud
        (HM99). A  red filament  is visible east of N88A.
        J, H and K are respectively coded B, G and R.  North is up and East is left. 
          }
 \label{fig:S54} 
\end{center}
\end{figure*}
\begin{figure*}[t]
\begin{center}
\hspace{0cm} 
\includegraphics[width=18.0cm]{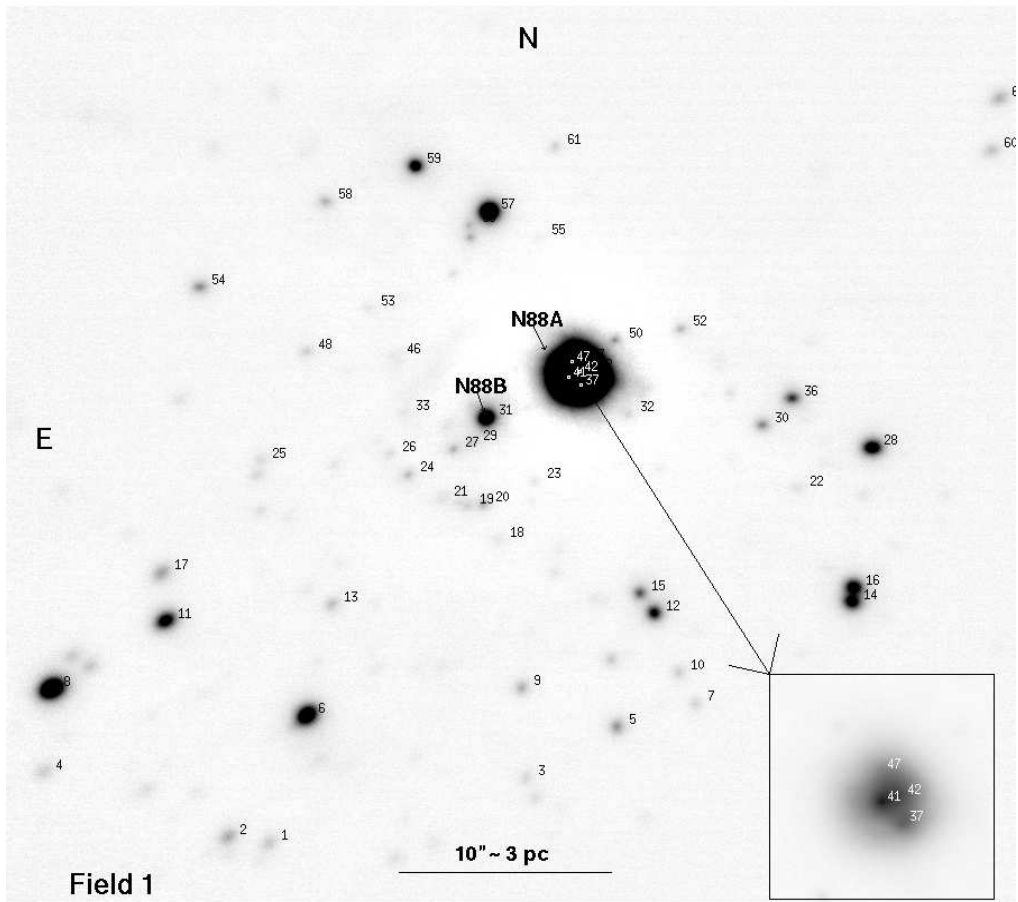}
                                     \vspace{-9cm}
\caption{Finding chart  in the J-band   obtained with the S54 camera (Field 1).  The inset
         corresponds to an enlargement (factor 2) and unsaturated image of  N88A extracted
         from Field 1. The numbering refers to Table 3. The total field size corresponds to
         48.6$\arcsec$ x 47.5$\arcsec$ (or $\sim$ 14 pc x 14 pc). } 
\label{fig:S54inset}
\end{center}
\end{figure*}
\begin{figure*}[t]
\begin{center}
\hspace{0cm} 
\includegraphics[width=18.0cm]{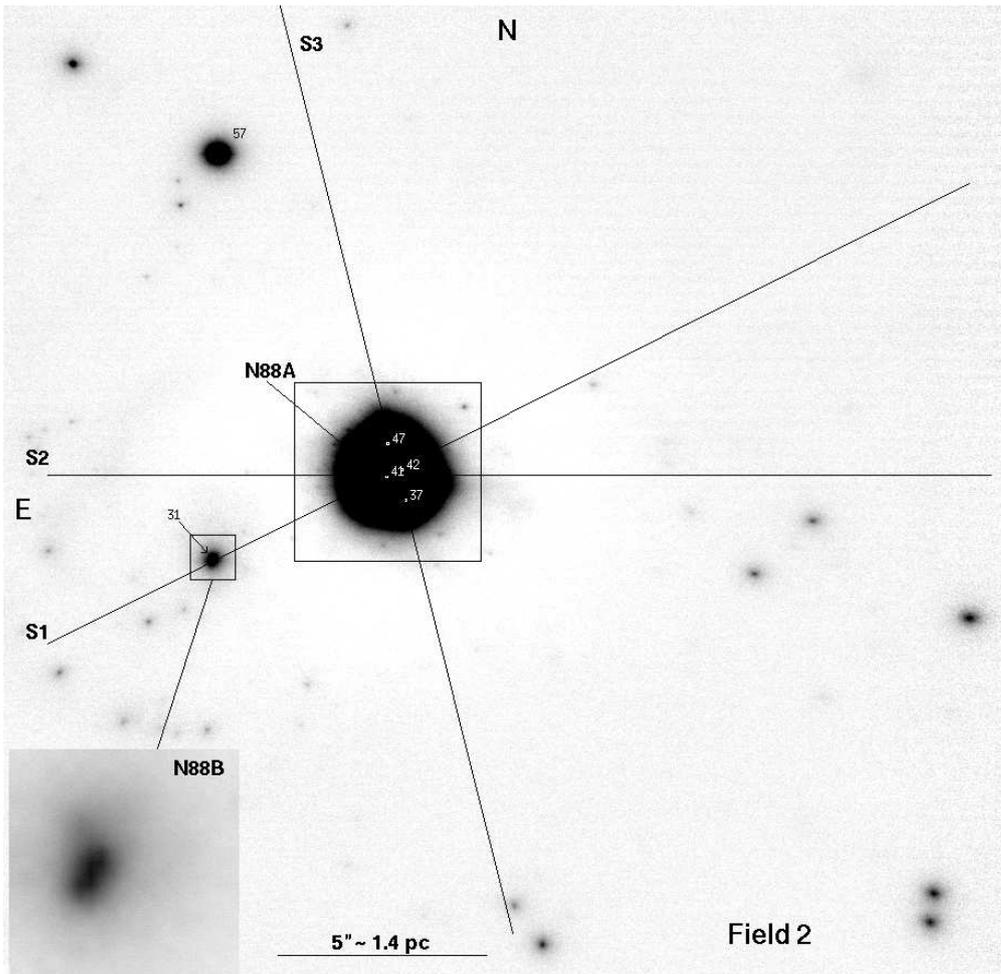}
                                       \vspace{-9cm}
\caption{Ks-band image of SMC-N88 obtained with the S27 camera
         (Field 2). The location of the slits labelled S1, S2 and S3  used in the spectroscopic
         mode is indicated by a solid line. In N88A the positions of the stars  $\#$37, $\#$41,
         $\#$42 and $\#$47 are indicated.
         The bottom left is an enlargement of 0.55$\arcsec$ x 0.55$\arcsec$ containing the complex object N88B.
         The image  size  corresponds to 25$\arcsec$ x 25$\arcsec$ (or $\sim$ 7 pc x 7 pc). }
\label{fig:S27inset}
\end{center}
\end{figure*}
\renewcommand{\arraystretch}{0.5}
\tabcolsep=0.04cm
\begin{table}
\caption{ Log of the photometric VLT/NACO observations.}
\begin{center}
\begin{tabular}{llccccc}
\noalign{\smallskip}
\hline
\hline
\noalign{\smallskip}
Id.       &  Filter  & Expo.   &Mode   &  Date      & Seeing    & FWHM  \\
          &          &t(s) x n &       &            &($\arcsec$)&($\arcsec$)      \\
\noalign{\smallskip}
\hline
\noalign{\smallskip}
N88A     &J 1.27 $\mu$m         &20 x 30   & -         &     -     & -         & 0.35     \\
         &H 1.66 $\mu$m         &20 x 30   & -         &     -     &           & 0.27     \\
         &Ks 2.18 $\mu$m        &30 x 30   & S54       &9/10/2004  &   0.6-0.9 & 0.21     \\
         &L' 3.8 $\mu$m         &0.2 x 26  & S27       &10/10/2004 &   0.9-1.2 & 0.10     \\
         &NB-2.12               &200x 20   & S54       &     -     &           & 0.19     \\
         &NB-2.24               &150x 20   & S54       &     -     &           & 0.25     \\
         &Ks                    &60 x 30   & S27       &11/10/2004 &           & 0.10     \\
\noalign{\smallskip}
\hline
\end{tabular}
\end{center}
\end{table}
\begin{figure*}[t]
\begin{center}
\hspace{-1cm} 
\includegraphics[width=18.0cm]{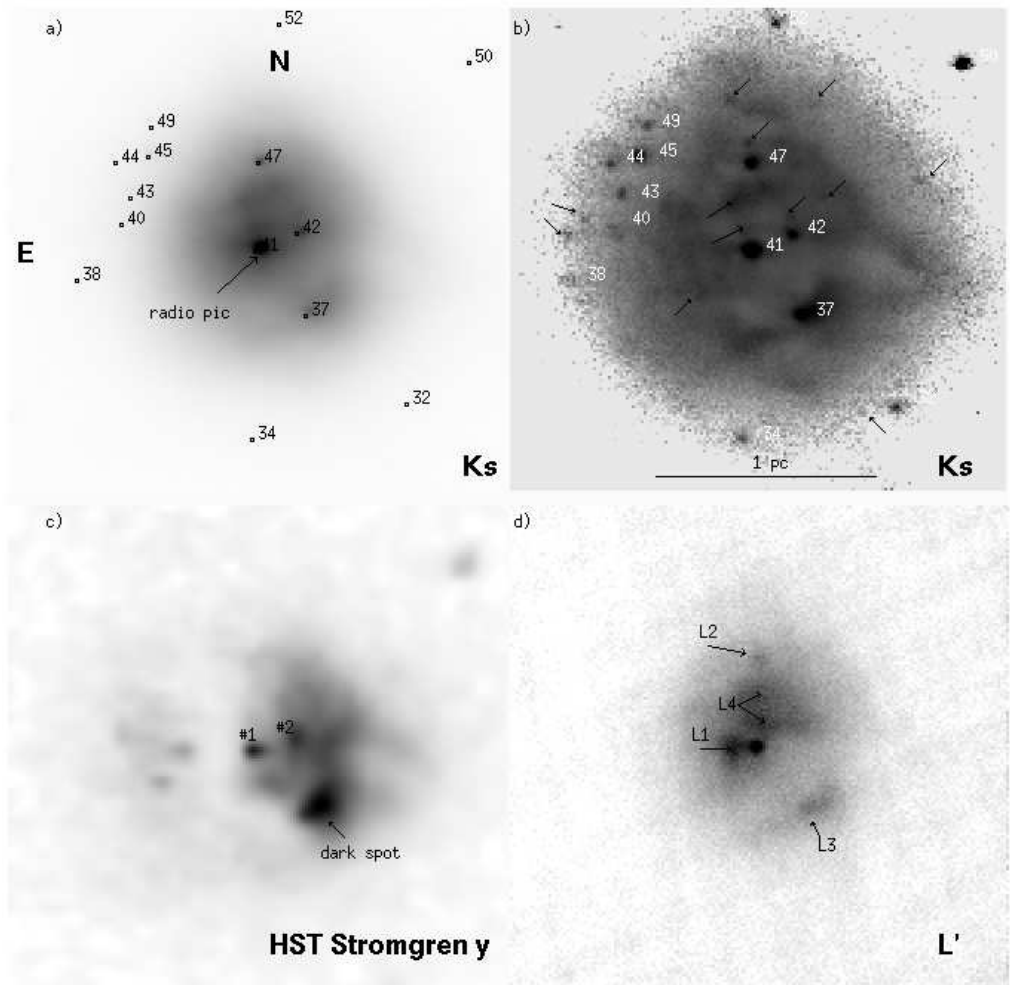}
                                            \vspace{-9cm}
\caption{Enlargements of the HEB N88A in Field 2. a) Ks-band obtained with the S27 camera.
        The numbering refers to  Table 4 (Field 2). b)  the DDP process  applied to 
        the Ks-band allows to  enhance  the faint stars  of N88A-cl. A few stars not identified with DAOPHOT
        and small features are indicated by arrows. c) Stromgren y image obtained with the HST showing the 
        absorption lane, the stars $\#$1, $\#$2 as well as  the `dark spot' of HM99. The Stars $\#$1
        and $\#$2 correspond to our stars  $\#$41 and  $\#$42.
        d) L'-band obtained with the S27 camera, the different components are indicated. The field 
        corresponds to 4.43$\arcsec$ x 4.32$\arcsec$ (or 1.27 pc x 1.27 pc) and is outlined in
        (Fig. \ref{fig:S27inset}). } 
 \label{fig:S27zoom}
\end{center}
\end{figure*}  
\renewcommand{\arraystretch}{0.5}
\tabcolsep=0.04cm
\begin{table}
\caption{ Log of the VLT/NACO long-slit spectroscopic observations.}
\begin{center}
\begin{tabular}{lcccccc}
\noalign{\smallskip}
\hline
\hline
\noalign{\smallskip}
Id.        &  Date      & Slit     &  Exposures & Mode     &$\lambda$/$\delta$$\lambda$& Seeing       \\
           &            & (mas)         &   t(s) x n &          &                           &($\arcsec$)   \\
\noalign{\smallskip}
\hline
\noalign{\smallskip}
S1         &9/10/2004  &172        & 100 x 10   &S54-4-SHK &   400                     & 0.6-0.9      \\
S2         &10/10/2004 & -         & -          &  -       &   -                       &  -           \\
S3         &11/10/2004 & -         & -          &  -       &   -                       &  -           \\
           &           &           &            &          &                           &              \\
hip8485    &9/10/2004  & -         & 2.5 x 4    &  -       &   -                       &   -          \\
hip29968   &9/10/2004  & -         & 2 x 4      &  -       &   -                       &   -          \\
hip103087  &11/10/2004 & -         & 1.8 x 4    &  -       &   -                       &   -          \\
\noalign{\smallskip}
\hline
\end{tabular}
\end{center}
\end{table}
\section{Observations and data reduction}
 NIR  observations of N88A were obtained at the ESO Very Large Telescope (VLT) during
 October 2004. Images and spectra were taken  using NACO on UT4, composed of the Nasmyth Adaptive
 Optics System (NAOS) and the High Resolution IR Camera and Spectrometer (CONICA).
 The detector was  a 1026 x 1024 SBRC InSb Alladin 3 array. The  cameras S54 and S27 were used
 in the range  1.0-2.5 $\mu$m and the L27 camera in the range 2.5-5.0 $\mu$m. 

 The field-of-view (FOV) of the S54 camera was  54$\arcsec$ x 54$\arcsec$  with a pixel size of
 0.05274$\arcsec$, corresponding to 0.015 pc at the distance modulus of 19.05 for the SMC
 (Kovacs 2000). The FOV  of the S27  and L27 camera was  27.15$\arcsec$ x 27.15$\arcsec$ with a
 pixel size of 0.02637$\arcsec$, corresponding to 0.0075 pc. For spectroscopy  we used  the S54 camera.
 
 As adaptive optics (AO) reference source for wavefront sensing we used  the object itself (N88A)
 (Testor et al. 2005). The conditions were photometric and the seeing  ranging from 0.65$\arcsec$
 to 1$\arcsec$ in the visible. After subtraction of the average dark frame, each image was divided by 
 the normalized flat field image. The data  were reduced mainly with the ESO software packages 
 MIDAS  and ECLIPSE.

\subsection{Imaging}
  Images  through J, H, Ks broad-band  and 2.12$\mu$m, 2.24$\mu$m  narrow band  filters were 
 obtained with the S54 camera. A composite JHKs colour image of the observed field
 is shown in Fig. \ref{fig:S54}. Images with higher spatial resolution, in the Ks and L'
 large band, were also obtained with the S27 camera. The log of NIR imaging observations
 is given in Table 1. The AutoJitter mode was used: that is, at each exposure, the  telescope
 moves according to a random  pattern in a 6$\arcsec$ x 6$\arcsec$ box. Table 1 lists the stellar
 full-width-at-half-maximum (FWHM) in final images of different observed bands of  the star at 
 J2000 coordinates ($\alpha$, $\delta$) = (1$^{h}$,24$^{m}$,8.88$^{s}$, -73$^{o}$,8',56.2") of 
 the 2MASS survey (Cutri et al. 2003). The AO image  is affected by anisoplanatism and leads
 to degradation of the point spread function (PSF) becoming more elongated  as the angular offset
 from the guide star increases. This   has been  taken  into consideration for the 
 photometric measurements, as explained in Sect. 2.3.

\subsection{Spectroscopy}
 Spectroscopy was performed  in the  S54-4-SHK mode (broad-band filter), giving a  linear dispersion
 of 1.94 nm/pixel and a spatial scale of 53 mas/pixel. Three long-slit spectra  S1, S2, and S3
 were chosen   from the  IR images given by NACO. S1 (PA = 115$\degr$) and 
 S2 (PA = 89.3$\degr$) cross the central bright star. S3 (PA = 18.9$\degr$) crosses the stars  $\#$37
 and  $\#$47  (see Fig. \ref{fig:S27inset}). The slit width was 172 mas and the spectral  resolution 
 $\sim$ 400. For each  exposure the detector integration time  (DIT) was 100s. Ten  exposures were
 obtained in the Autonod on slit mode, which allows spectroscopy of moderately extended objects.
 The log of spectroscopic observations is given in Table 2. In order to  remove telluric absorption
 features, stars with a similar airmass were observed as  telluric standards. Spectroscopy was reduced
 with the MIDAS software package LONG. 
\tabcolsep=0.1cm
\begin{table*}[t]
\caption{ Stars in the region of N88A observed with the S54 camera (Field 1).}
\begin{center}
\begin{tabular}{llllllrrr}
\noalign{\smallskip}
\hline
\hline
\noalign{\smallskip}
Id.  & \hspace{0.2cm} $\alpha$(2000) &\hspace{0.2cm}$\delta$(2000)&mag-J&mag-H&Mag-K&J-K&J-H&H-K\\
\noalign{\smallskip}
\hline
\noalign{\smallskip}
              1    & 1  24  11.42  & 73   9  26.15&   18.24&  17.65&  17.50&   0.74&   0.59&   0.15\\
              2    & 1  24  11.86  & 73   9  25.85&  17.63&  17.71&  17.65&  -0.02&  -0.08&   0.06\\
              3    & 1  24   8.67  & 73   9  23.08&  19.08&  18.88&  18.84&   0.24&   0.20&   0.04\\
              4   & 1  24  13.84  & 73   9  22.77&  18.46&  18.48&  18.45&   0.01&  -0.02&   0.03\\
              5   & 1  24   7.69  & 73   9  20.63&  17.89&  17.61&  17.34&   0.55&   0.28&   0.27\\
              6   & 1  24  11.02  & 73   9  20.11&  15.76&  15.85&  15.82&  -0.06&  -0.09&   0.03\\
              7   & 1  24   6.85  & 73   9  19.56&  19.16&  18.71&  18.51&   0.66&   0.45&   0.21\\
              8   & 1  24  13.75  & 73   9  18.85&  15.17&  15.37&  15.38&  -0.21&  -0.20&  -0.01\\
              9   & 1  24   8.72  & 73   9  18.79&  18.22&  18.18&  18.02&   0.20&   0.04&   0.16\\
              10   & 1  24   7.03  & 73   9  18.05&  18.80&  18.79&  18.71&  0.09&    0.01&  0.08\\
              11   & 1  24  12.53  & 73   9  15.61&   16.27&  16.45&  16.47&  -0.20&  -0.18&  -0.02\\
              12   & 1  24   7.30  & 73   9  15.25&  16.72&  16.87&  16.87&  -0.16&  -0.15&  -0.00\\
              13   & 1  24  10.75  & 73   9  14.84&  18.67&  18.53&  18.06&   0.61&   0.14&   0.47\\
              14   & 1  24   5.18  & 73   9  14.70&  16.27&  16.44&  16.31&  -0.04&  -0.17&   0.13\\
              15   & 1  24   7.45  & 73   9  14.32&  17.31&  17.48&  17.47&  -0.15&  -0.17&   0.01\\
              16   & 1  24   5.16  & 73   9  14.07&  16.19&  16.40&  16.26&  -0.07&  -0.21&   0.14\\
              17   & 1  24  12.57  & 73   9  13.33&  17.44&  17.59&  17.40&   0.04&  -0.15&   0.19\\
              18   & 1  24   8.97  & 73   9  11.82&  19.78&  19.65&  19.52&   0.26&   0.12&   0.14\\
              19   & 1  24   9.30  & 73   9  10.17&  18.44&  18.21&  17.87&   0.57&   0.23&   0.34\\
              20   & 1  24   9.13  & 73   9  10.09&  18.30&  18.23&  18.19&   0.11&   0.08&   0.03\\
              21   & 1  24   9.58  & 73   9   9.84&  19.11&  18.51&  17.66&   1.45&   0.60&   0.84\\
              22   & 1  24   5.76  & 73   9   9.34&  18.89&  18.59&  17.93&   0.96&   0.30&   0.66\\
              23   & 1  24   8.58  & 73   9   9.01&  19.65&  19.34&  18.95&   0.70&   0.31&   0.39\\
              24   & 1  24   9.93  & 73   9   8.68&  18.66&  18.36&  17.80&   0.86&   0.31&   0.55\\
              25   & 1  24  11.52  & 73   9   8.03&  18.89&  18.73&  18.51&   0.38&   0.15&   0.22\\
              26   & 1  24  10.13  & 73   9   7.70&  19.48&  19.46&  19.30&   0.17&   0.02&   0.15\\
              27   & 1  24   9.45  & 73   9   7.50&  18.37&  18.18&  17.97&   0.40&   0.19&   0.20\\
              28   & 1  24   4.96  & 73   9   7.42&  16.32&  16.48&  16.42&  -0.10&  -0.16&   0.06\\
              29   & 1  24   9.25  & 73   9   7.17&  18.21&  18.11&  17.88&   0.33&   0.10&   0.24\\
              30   & 1  24   6.14  & 73   9   6.35&  17.85&  17.51&  17.33&   0.52&   0.34&   0.18\\
              31   & 1  24   9.09  & 73   9   5.99&  15.74&  15.73&  15.51&   0.23&   0.01&   0.16\\
              32   & 1  24   7.57  & 73   9   5.80&  19.04&  19.09&  18.91&   0.13&   -0.05&  0.18\\
              33   & 1  24   9.98  & 73   9   5.77&  19.46&  18.76&  17.83&   1.63&   0.70&   0.93\\
              36   & 1  24   5.82  & 73    9   5.09&  17.36&  17.43&  17.26&  0.10&   -0.07&  0.17\\
                     &               &              &       &       &       &       &       &        \\
             37$^{n}$   & 1  24   7.84  & 73   9   4.24&  15.01&  15.10&  14.28&   0.73&  -0.09&   0.81\\   
             41$^{n}$   & 1  24   7.96  & 73   9   3.74&  14.68&  14.56&  13.82&   0.86&   0.12&   0.74\\    
             42$^{n}$   & 1  24   7.85  & 73   9   3.60&  15.05&  15.05&  14.44&   0.61&   0.00&   0.61\\    
             47$^{n}$   & 1  24   7.96  & 73   9   3.00&  15.55&  15.36&  14.88&   0.67&   0.19&   0.48\\    
                 &               &              &       &       &       &       &       &        \\
             46   & 1  24  10.08  & 73   9   3.05&  19.65&  18.93&  18.18&   1.47&   0.72&   0.75\\
             48   & 1  24  11.02  & 73   9   2.86&  18.93&  18.90&  18.82&   0.12&   0.03&   0.09\\

       50$^{n}$   & 1  24   7.53  & 73   9   2.09&  17.61&  17.62&  17.16&   0.45&  -0.01&   0.46\\   
                                
             52   & 1  24   7.01  & 73   9   1.79&  18.80&  18.98&  18.57&   0.23&  -0.18&   0.41\\
             53   & 1  24  10.35  & 73   9   0.80&  19.91&  19.51&  18.95&   0.96&   0.40&   0.56\\
             54   & 1  24  12.17  & 73   8  59.81&  17.75&  17.59&  17.29&   0.46&   0.16&   0.30\\
             55   & 1  24   8.53  & 73   8  57.45&  20.05&  20.01&  19.65&   0.40&   0.04&   0.36\\
             56   & 1  24   9.27  & 73   8  56.87&  17.57&  17.42&  17.29&   0.28&   0.15&   0.13\\
             57   & 1  24   9.07  & 73   8  56.24&  15.36&  14.88&  14.59&   0.77&   0.48&   0.29\\
             58   & 1  24  10.81  & 73   8  55.78&  18.45&  18.43&  18.26&   0.19&   0.02&   0.17\\
             59   & 1  24   9.86  & 73   8  54.07&  16.51&  16.67&  16.64&  -0.13&  -0.16&   0.03\\
             60   & 1  24   3.68  & 73   8  53.33&  18.51&  18.03&  17.59&   0.92&   0.48&   0.44\\
             61   & 1  24   8.36  & 73   8  53.17&  19.27&  18.65&  18.44&   0.83&   0.62&   0.21\\
             62   & 1  24   3.59  & 73   8  50.89&  18.06&  17.65&  17.43&   0.63&   0.41&   0.22\\

\noalign{\smallskip}
\hline
\end{tabular}
\end{center}
 $^{n}$ Stars in N88A analyzed using the  NSTAR routine.
\end{table*}
\begin{table*}
\caption{Stars in  N88A observed with the S27 camera (Field 2) in the Ks-band (numbers refer to Fig. \ref{fig:S27zoom}.}
\begin{center}
\begin{tabular}{llllllllll}
\noalign{\smallskip}
\hline
\hline
\noalign{\smallskip}
Id.       &&\hspace{0.2cm} $\alpha$(2000)&\hspace{0.2cm}$\delta$(2000)&K$^{dao}$&  err &K$^{psf}$ &L$^{ap}$&K-L \\
\noalign{\smallskip}
\hline
\noalign{\smallskip}
57         &&  1   24   8.88 &  -73    8  56.00&  14.64   &      0.003&               &14.36$\pm$0.05&       \\
34         &&  1   24   7.94 &  -73    9  5.69 &  18.45   &      0.070&               &              &       \\
35         &&  1   24   7.64 &  -73    9  5.42 &  18.73   &      0.050&               &              &       \\
37         &&  1   24   7.83 &  -73    9   4.56&  16.43   &      0.032& 16.66$\pm$0.15&              &       \\
38         &&  1   24   8.30 &  -73    9   3.85&  18.46   &      0.060&               &              &       \\
40         &&  1   24   8.21 &  -73    9   3.80&  18.40   &      0.065&               &              &       \\
41         &&  1   24   7.93 &  -73    9   3.99&  14.99   &      0.010& 15.05$\pm$0.10&14.1$\pm$0.10 &  0.95 \\
42         &&  1   24   7.85 &  -73    9   3.82&  16.11   &      0.047& 16.60$\pm$0.30&              &       \\
43         &&  1   24   8.21 &  -73    9   3.30&  17.62   &      0.066&               &              &       \\
44         &&  1   24   8.23 &  -73    9   3.05&  17.76   &      0.060&               &              &       \\
45         &&  1   24   8.16 &  -73    9   3.14&  17.12   &      0.044&               &              &       \\
47         &&  1   24   7.93 &  -73    9   3.19&  15.62   &      0.036& 15.99$\pm$0.20&              &       \\
49         &&  1   24   8.16 &  -73    9   2.70&  17.41   &      0.061&               &              &       \\
51         &&  1   24   7.49 &  -73    9   2.29&  18.27   &      0.040&               &              &       \\
52         &&  1   24   7.88 &  -73    9   1.90&  18.65   &      0.061&               &              &       \\
     L1-C  &&  1   24   7.95 &  -73    9   3.80&$\geq$18.5&      0.024&               &13.98$\pm$0.20& $\geq$4.52      \\
\noalign{\smallskip}
\hline
\end{tabular}
\end{center}

$^{dao}$, $^{psf}$ and $^{ap}$ magnitudes derived using DAOPHOT, the PSF of star $\#$57, and an aperture respectively.  
\end{table*}
\subsection{Photometry }
 In  Fig. \ref{fig:S54inset}  we   present the N88 region  observed with the S54 camera (Field 1)
 through  the J-band filter.
 The instrumental magnitudes of the elongated  stars (see Sect. 2.1)  outside the central region,
 were derived with
 DAOPHOT (Stetson 1987), using concentric aperture photometry  to integrate all the flux
 of each  star. Although PSF photometry is   better adapted for crowded fields,
 we could not use it.  Indeed  the stars in our  field 
 were too faint and/or crowded to obtain the number of   PSF stars necessary  to use the
 photometric analysis elaborated by  Pugliese et al (2002)  taking into account the
 AO anisoplanatism effect.

 The detected stars are identified by a number referring to Column 1 of Table 3 that gives the 
 astrometry and photometry. The central object N88A is not affected by anisoplanatism,
 so the JHK instrumental magnitudes of the stars were derived using the  DAOPHOT's
 multiple-simultaneous-profile-fitting photometry routine (NSTAR),  well adapted for
 photometry in crowded fields. The detected stars are shown in the inset of
 Fig. \ref{fig:S54inset} and are also  listed in Table 3. Almost all the stars of Field 1
 (Fig. \ref{fig:S54inset}) have photometric  uncertainties in the J-, H- and K-band,
 less than 0.03 mag for stars with K $<$ 16 mag, less than 0.06 for stars with 16 $<$ K $<$ 18
 mag and greater than 0.1 for stars 18 $<$ K $<$ 20 mag.

 Fig. \ref{fig:S27inset} (Field 2)  shows  only a part of N88 observed with the S27 camera
 through the Ks-band filter. On this  figure the directions of the spectra S1, S2 and S3 are
 plotted. This camera, with a pixel two times smaller than  S54, has a  
 better spatial resolution, so the analysis of the crowded field of N88A with the routine NSTAR
 allows  a  more accurate star detection. In Fig. \ref{fig:S27zoom}a the stars of N88A are 
 identified by a number referring to Column 1 of Table 4. In this table, the  average photometric errors 
 of  the  stars  reported by DAOPHOT are $\sim$0.04 mag in J and 0.06 in H and Ks for stars of
 magnitude $\leq$ 16 except for the bright isolated  star $\#$57 outside N88A ($\sim$0.03 
 and $\sim$0.05 mag). This star will be used as PSF.
\begin{figure}
\begin{center} 
\includegraphics[width=6cm,height=9cm,angle=-90]{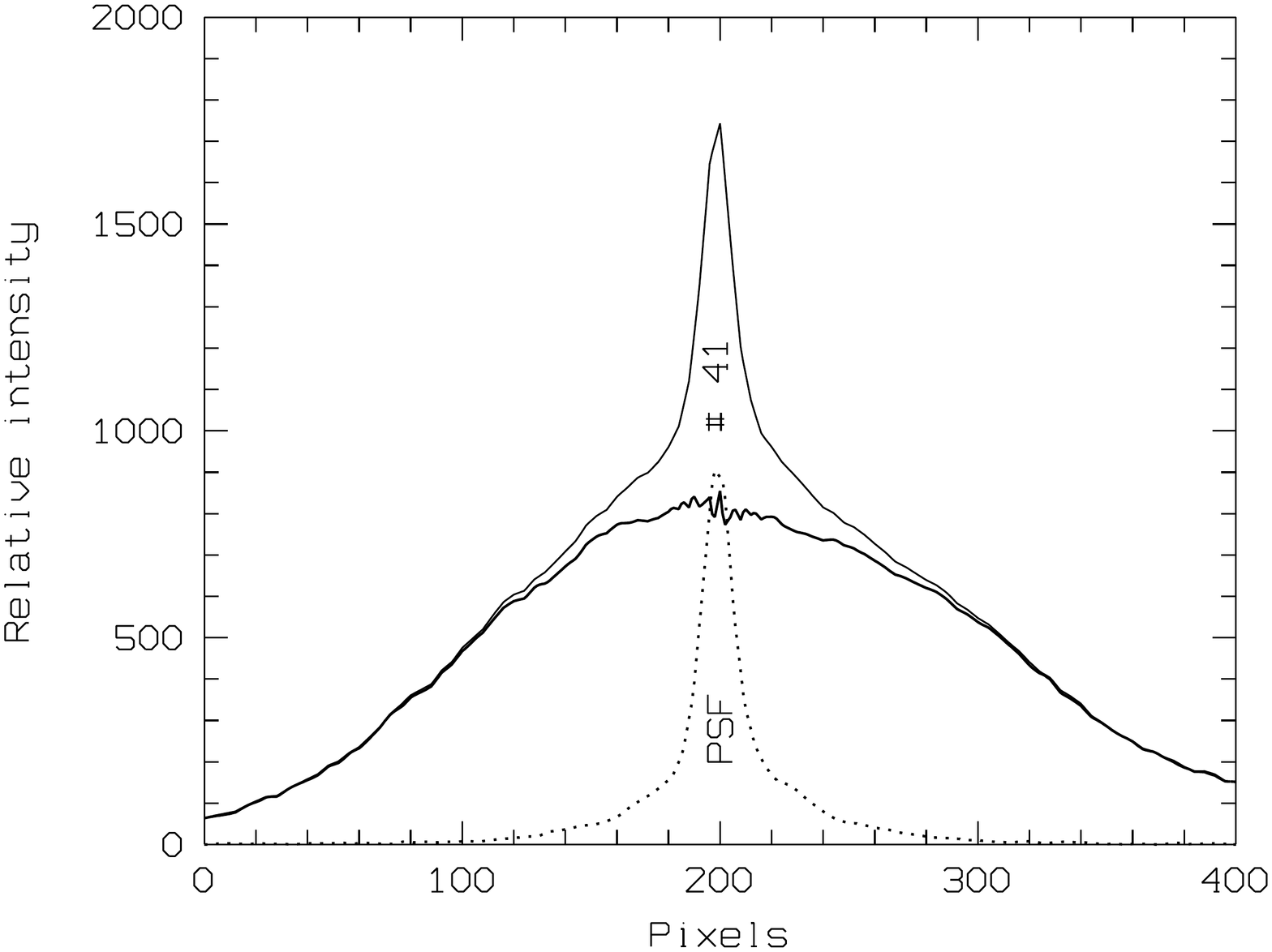}
\caption{The thin solid lines show  the profile crossing the center 
        of star $\#$41 without subtraction of the continuum. The dotted line shows the profile (PSF)  of
        the isolated  star $\#$40 labelled $\#$80 in Field 1.
        The intensity of the PSF  of FWHM = 0.11$\arcsec$ is multiplied by a normalization factor.
        The thick solid line shows the continuum  obtained after the subtraction of the PSF. The plot range
        corresponds to 2.64$\arcsec$.}
 \label{fig:PSF}
\end{center}
\end{figure}
 The photometric calibration was obtained using  the isolated 2MASS star at J2000
 ($\alpha$, $\delta$) = (1$^{h}$,24$^{m}$,8.88$^{s}$, -73$^{o}$,8',56.2")
  corresponding  to our star $\#$57. The conversion of pixel coordinates to
 $\alpha$ and $\delta$  was derived  using the same  star and  the relative positions
 of our stars are accurate to better than 0.1$\arcsec$.

 In the core of N88A  the  determination of the  sky aperture parameters used in NSTAR
 is very sensitive, even  with the S27 camera. Indeed, the wings of the stars  superpose
 the wings of the  strong  continuum. The distribution of this continuum  resembles
 a gaussian profile (Fig. \ref{fig:PSF}).
 The error on the magnitude of these stars, due to a steeply sloping continuum background, is
 greater than the error given by DAOPHOT.  Because of this situation, the K magnitude of
 the central  star labelled  $\#$41
  at low  and  at higher spatial resolution  (Field 1 and Field 2) is respectively 13.82  and 14.99 mag.
 Therefore, the K magnitudes of the stars  $\#$37, $\#$41, and $\#$47 (Field 2)
 were remeasured   by subtracting a  one-dimensional (1-D) profile corresponding to the PSF
  crossing the center of the
 isolated reference  star $\#$ 57. The magnitude of the PSF  was multiplied
  by a factor  in such 
 a way that  only the continuum remains visible. In this case its  magnitude  corresponds
 to the magnitude of the  star.
 An example is given in Fig. \ref{fig:PSF}. 
 The K magnitudes of the stars obtained  with this method are listed in col. 7 of Table 4.
 In this table  the magnitude of  $\#$41 is in agreement with
 the magnitude obtained with NSTAR (col 5), while
 for  $\#$37, $\#$42 and $\#$47  the Ks magnitude is greater. Several faint stars
 under the detection  level, or slightly extended, are not detected or rejected  by DAOPHOT
 (Fig. \ref{fig:S27zoom}b).
\begin{figure}
\begin{center}
\hspace{0cm}
\includegraphics[width=6cm,height=9cm,angle=-90]{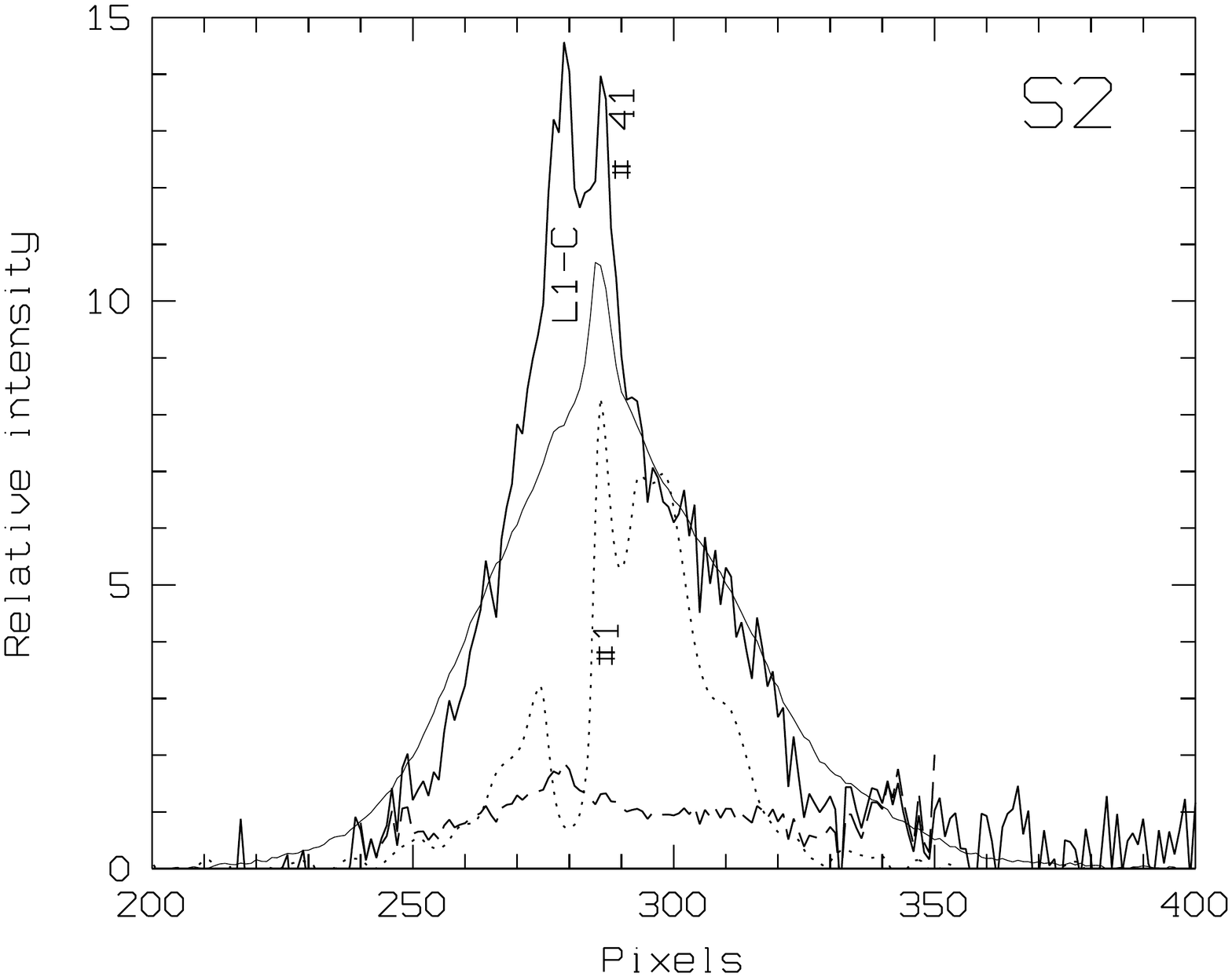}
\caption{East-west intensity distributions along the direction of slit S2  extracted
        from the Ks image (thin line),  L' image (thick line) and Stromgren y (HST) images (dotted line).
        The  L'/Ks ratio (dashed line) shows  that the  core L1-C (see Sec. 3.3.1) coincides with  the
        minimum  intensity of the absorption lane on the  Stromgren y plot. The plot range corresponds 
        to 5.1$\arcsec$.}
\label{fig:KS_L}
\end{center}
\end{figure} 
 In our underexposed  L'-band frame of N88, not shown here, contrary to the JHK-band,
 no star is found except star $\#$57. However, in N88A the  bright 
 star  $\#$41 is  visible as well as the  stars  $\#$37, $\#$42, $\#$47 and several
 unresolved features. A peculiar bright core labelled L1-C (see Sect. 3. below) located 0.2$\arcsec$ east
 of  $\#$41 is also found. This core coincides approximatively with the HST absorption  lane  (HM99)
 and has a very faint counterpart  in the Ks-band (Fig. \ref{fig:KS_L}). In order to derive
 the  L' magnitude of these objects, we  referred to  the  L' photometry of
 Israel $\&$ Koorneef (1991), obtained through  a 7$\arcsec$ aperture (Table 5). The 
 integration of N88A  in  our sky subtracted  Ks image,  using a  4$\arcsec$  aperture,
 gives a magnitude of 11.08,  in agreement with Israel $\&$ Koorneef (1991) 
 and other authors (Table 5). In an aperture of 4$\arcsec$  we have integrated the L' flux 
 of N88A that was  then calibrated with L' = 8.92 mag given by Israel $\&$ Koorneef (1991). 
 The L' magnitudes of the stars  $\#$57, $\#$41 and the core  L1-C (continuum subtracted) were
 obtained using  an  aperture of 0.35$\arcsec$ and are listed in Table 4. 
                            \renewcommand{\arraystretch}{0.3}
                            \tabcolsep=0.005cm
\begin{figure}
\hspace{-0.1cm}
\includegraphics[width=6cm,height=8.5cm,angle=-90]{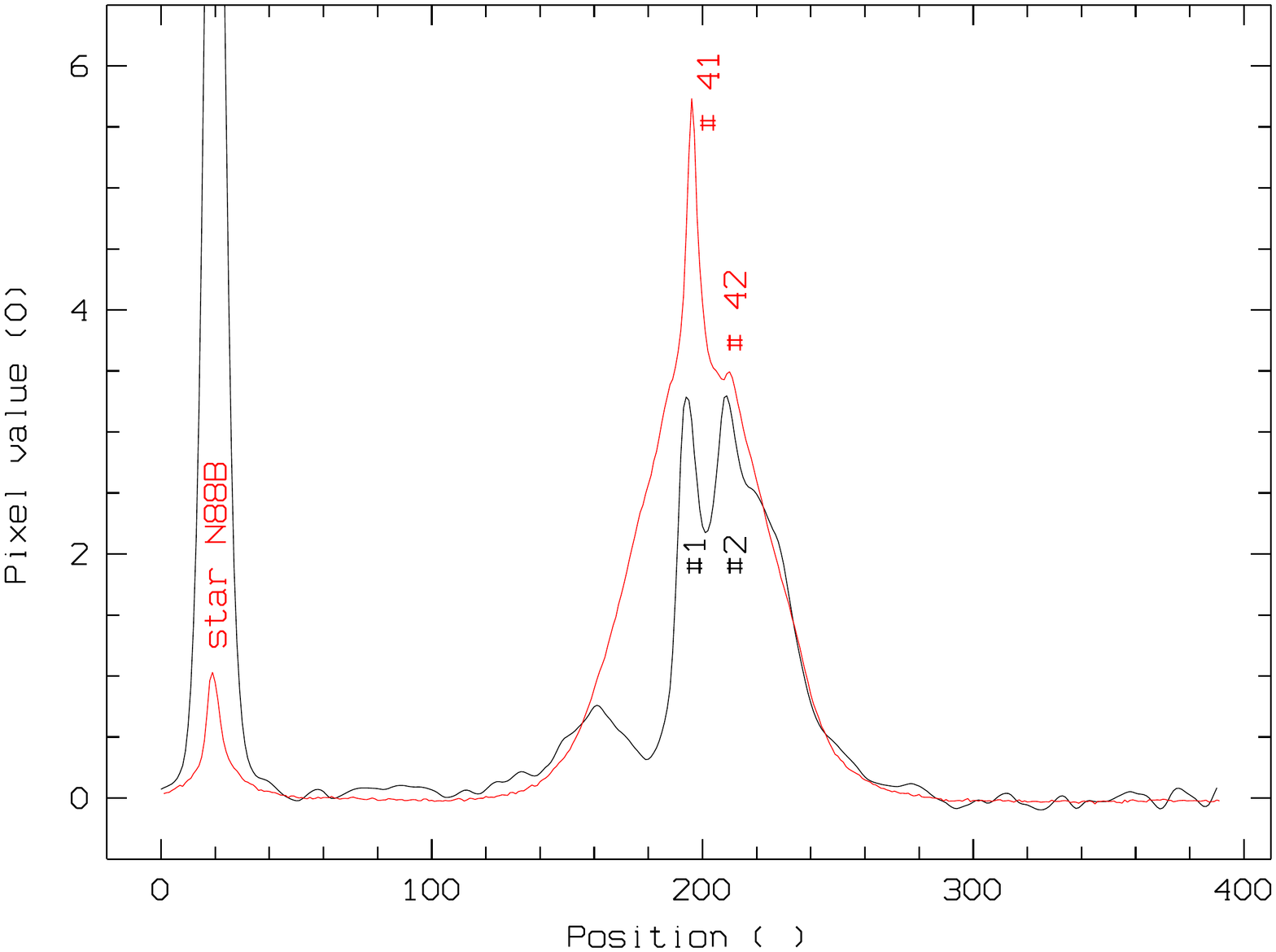}
\caption{Intensity distributions along the direction of slit S1 extracted from the Ks-band image 
        (red line) and HST Stromgren y  images (black line). The Stars $\#$1, $\#$2 and the
        absorption lane (HM99) as well as our star $\#$41 and  $\#$42
        are indicated.  The plot range corresponds to 10.6$\arcsec$. }
\label{fig:supKsHST}
\end{figure}
\section{Results and discussion}
\subsection{The HEB N88A}
  The HST  H$\alpha$  (F656N) and  continuum Stromgren y (F547M)
  images of N88A  described in HM99, show   two inhomogeneous wings  separated by a north-south
 absorption lane (Fig. \ref{fig:S27zoom}c). The  western  wing  is much brighter and
 contains two faint stars, $\#$1 and $\#$2, corresponding to our stars  $\#$41 and
 $\#$42, as well as a `dark spot' to the south at the location of our star $\#$37.
 Fig. \ref{fig:supKsHST} shows  the intensity distribution plots  in the Stromgren
 y-  and Ks-band  in the direction of the slit S1 (Fig. \ref{fig:S27inset}) crossing
  N88B and the star $\#$41. In  Fig. \ref{fig:supKsHST} the stars  $\#$41 and  $\#$42 
 of magnitude 14.99 and 16.11  respectively (Table 4) coincide with the  two faint  stars $\#1$
 and $\#$2  (Fig. \ref{fig:S27zoom}c) of y = 18.2 mag and 18.3 mag (HM99).

 Through the  Ks filter (Figs. \ref{fig:S27zoom}a and b), N88A appears as a circular nebular
 region of  $\sim$ 3.4$\arcsec$ diameter centered on the  relatively bright star  $\#$41.
 This star coincides with  the  2MASS point source 012407.92-730904.1 of K = 11.18 mag  
 (Cutri et al. 2003). In  a diameter  of $\sim$ 3.6$\arcsec$ our  N88A image  exhibits
 a small embedded cluster labelled N88A-cl of at least thirteen stars (Fig. \ref{fig:S27zoom}b).
 In Figure \ref{fig:S27zoom}b the  usual Digital Development Process (DDP)
 introduced by Okano was applied
  to enhance the faint stars by compressing the range of brightnesses between the bright
 and dim portions of the image. The K photometry  of these stars is listed in Table 4.
 These stars mainly concentred to the north and east, superpose  numerous nebular structures.
 Interestingly these stars are aligned in the direction of the interface between the HII regions
 N88A and N88B (HM99).Through the L'-band, N88A seems to be essentially formed by four distinct components 
 labelled L1, L2, L3 and L4 (Fig. \ref{fig:S27zoom}d). L1, that contains the core labelled
 L1-C (Fig. \ref{fig:supKsHST}), is the brightest and most compact component. In the Ks-band L1-C
 (Figs. \ref{fig:S27zoom}a and \ref{fig:KS_L}) shows a very faint counterpart 
 (L-Ks $\geq$ 4.2). L2 and L3 are  more diffuse and  coincide with the stars  $\#$47
 and $\#$37. L4 is bright, extended and formed by two east-west elongated
 subcomponents spanning  between stars  $\#$47 and  $\#$41.   
 In the L'-band the star  $\#$41 is relatively bright (L' $\sim$ 14.1 mag) and well resolved
 (Fig. \ref{fig:S27zoom}d). All these components superpose a diffuse nebular  continuum.
  On  the  y (F547M)  continuum  image (Fig. \ref{fig:S27zoom}c)
  the  `dark spot' corresponding to our component L3  is  very bright. 
 The y continuum structures  located between stars  $\#$41 and  $\#$37, as well as 
 north to the `dark spot' (Fig. \ref{fig:S27zoom}c), are not seen  in the L'-band (Fig. \ref{fig:S27zoom}d).
\subsection{The N88B region}
 At the optical wavelengths, HM99 found that the central star of N88B, corresponding 
 to our star  $\#$31, has an integrated magnitude  of y  = 16.57 and  consists
 of  at least three components. Our high spatial resolution  Ks-band image also shows
 that star $\#$31 of magnitude K = 15.74  and J-K = 0.23 is complex and formed  of  at least
 three components visible in the inset of Fig. \ref{fig:S27inset}. Two of them 
 oriented approximatively  south-north are relatively bright, whereas the one to the north-east
 corresponds  to a  faint diffuse feature. To the east of N88B  lies a red bow-shaped
 filament  with a  curvature radius of $\sim$ 3$\arcsec$  (Fig. \ref{fig:S54}) centered
 on N88B (see Sect. 3.6). This filament coincides with the narrow filament north-east of
 component B detected  in the  H$\alpha$-band (HM99).
\begin{figure}
\begin{center}
\hspace{-1cm}
\includegraphics[width=8cm,height=9.9cm,angle=-90]{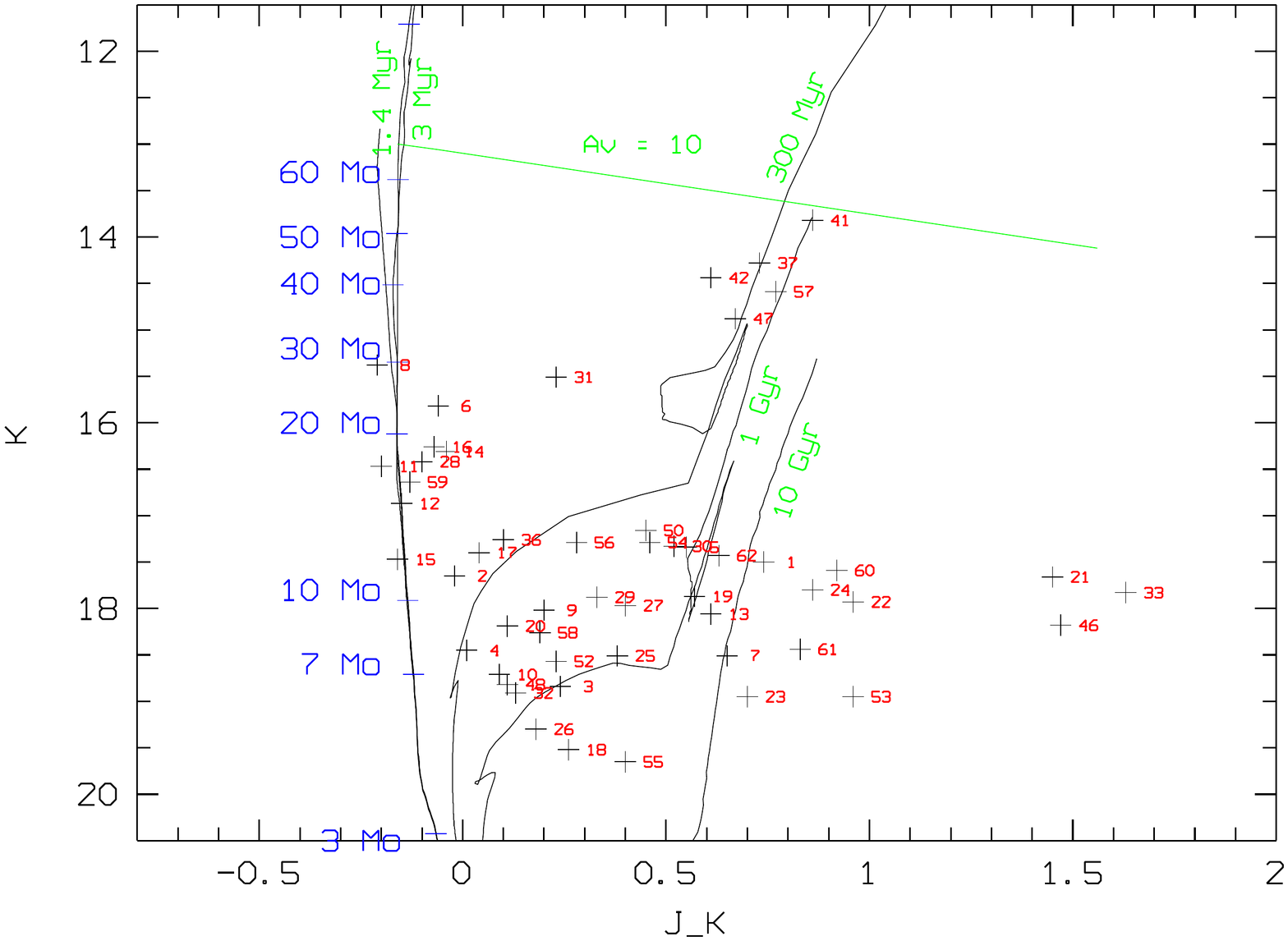}
\caption{ J-K CM diagram for the point sources measured in N88 and N88A (Field 1). From  left
        to right are overplotted the 1.4 Myr, 3 Myr, 300 Myr, 1 and 10 Gyr isochrones (extinction free).
        Some masses between 3$M_{\sun}$  and 60$M_{\sun}$ are marked with a tick on the 3 Myr isochrone.
        The stars  $\#$37,  $\#$41, $\#$42 and  $\#$47 of Field 2 (Table 4)
         analyzed using NSTAR (crosses) and the PSF of star $\#$57 (filled squares) are overplotted.}
\label{fig:HR}
\end{center}
\end{figure}
\begin{figure}
\begin{center}
\hspace{-0.7cm}
\includegraphics[width=8cm,height=9.5cm,angle=-90]{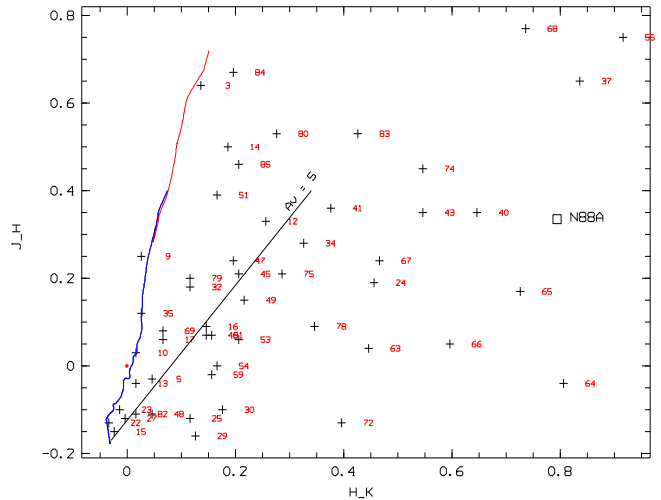}
\caption{CC diagram (J-H vs. H-K) for the measured  stars in N88 and N88A.
        The blue line is the main sequence for stars of age 1Myr and the red line for stars of age 10 Myr.  
        The empty square  represent the  colour-colour of N88A integrated in an aperture of 4$\arcsec$ diameter
 and the filled squares the bright stars in N88A-cl.  
        The solid line is the reddening vector up to Av = 5.}
\label{fig:HKJH}
\end{center}
\end{figure} 

\subsection{JHK CM and CC  diagrams}

 Fig. \ref{fig:HR} shows a Ks vs. J-K CM diagram  of the N88 region.
 The magnitudes of the stars (Field 1) belonging to N88A are systematically
 underestimated (see Sect. 2.3).  This difference is visible on the diagram  where  the stars of
 N88A (Field 2)  analysed using the PSF of star $\#$57 (Table 4) are overplotted.
 However, taking into account the PSF given by DAOPHOT for JHK of about 0.2-0.3$\arcsec$ we assume that
 the  J-K colour values  as well as J-H and H-K  are  correct within the uncertainties.
The color excess E(B-V) towards  N88 derived for hot stars  from the Magellanic Clouds Photometric
 Survey (Zaritsky et al. 2002) in a radius of  1 $\arcmin$,  is small  $\sim$  0.15.
 On Fig.  \ref{fig:HR} the  reddening track for O stars  is plotted, assuming a total visual
 extinction Av = 5.8E(J-K) (Tapia et al. 2003) and Ak = 0.112Av (Rieke $\&$ Lebosky 1985).
 It was derived using as reference  the star $\#$8 of type O9.5 in Wilcots 1994b.
 For this star we adopted a (J-K)$_{0}$ of -0.15 mag (Lejeune $\&$ Shaerer 2001).

  Several isochrones with different ages corresponding to  Z = 0.004  are overplotted (Fig.  \ref{fig:HR}).
 The diagram appears to reveal two populations. The first one is a young population of dwarf
 and  massive O stars which appears to be fitted with the  3 Myr isochrone.
 The second one  could be  a clump of red giant stars of  K magnitude in the range
 of 17-19.5 expanding in the age 300 Myr-10 Gyr. The stars lying beyond  the 10 Gyr isochrones
 are likely to represent embedded stars  situated deeper in the molecular cloud, young stars with
 circumstellar material or evolved stars surrounded by dust.

 Fig. \ref{fig:HKJH} shows the  H-K vs. J-H CC diagram. In this figure  the solid line 
 represents the reddening vector up to Av = 5 mag.  All stars that lie on the right side of the
 reddening vector should have IR-excess. Due to uncertainties on H-K colour we  take into account
 only the stars beyond $\sim$ 0.1 mag  to the right of the reddening vector. Hence, the number of stars with
 IR-excess extracted from  the  CC diagram corresponds to at least  30$\%$ of  the measured stars.
 On the CC diagram, obtained after integration in an  aperture  of 4$\arcsec$ diameter (Table 5), N88A is  overplotted 
 and is found  to the extreme right (Fig. \ref{fig:HKJH}). The plot shows a red  J-H colour
 of 0.33 mag which contrasts with  the blue  J-H colour given by  Israel $\&$ Koorneef (1991).   
                                                                                                                                                                                                                                                                                                              
\renewcommand{\arraystretch}{0.5}
\tabcolsep=0.08cm
\begin{table}
\caption{ Integrated NIR Photometry of the whole region  N88A}
\begin{center}
\begin{tabular}{lcccccccl}
\noalign{\smallskip}
\hline
\hline
\noalign{\smallskip}
 Reference         &  J      &  H           &  K          &  L'    & J-K   & Ks-L'     & FWHM         &Av    \\
\noalign{\smallskip}
                   &         &              &             &        &       &           & ($\arcsec$)     &       \\
\noalign{\smallskip}
\hline
\noalign{\smallskip}
  Denis$^{d}$      &  12.19  &              &11.11        &        &1.08  &            & 4            & 7.3  \\
  2MASS$^{m}$      &  12.31  &11.98         &11.18        &        &1.13  &            & 4            & 7.3  \\
  Israel$^{i}$     &         &              &11.05        &8.92    &      &2           & 10           &      \\
 this paper        & 12.15   &11.92         &11.08        &        &      &            & 4            &  7.1 \\
                   &         &              &             &        &      &            &              &       \\ 
\noalign{\smallskip}
\hline
\end{tabular}
\end{center}
 $^{d}$ Cioni et al. (2000), $^{m}$ Cutri et al. (2003), $^{i}$ Israel $\&$ Koorneef (1991).
  From  the Infrared Array Camera (IRAC) archive,  Charmandaris et al. (2008) using an aperture 
  of $\sim$7$\arcsec$, give for the wavelength
 3.6$\mu$m (L-band), 4.5$\mu$m, 5.8$\mu$m and  8$\mu$m a magnitude for N88A of 9.52, 8.29, 7.1 and
 5.58 mag respectively.
\end{table}
\begin{figure}[t]
\begin{center}
\hspace{0cm}
\vspace{0cm}
\includegraphics[width=8cm,height=9cm,angle=-90]{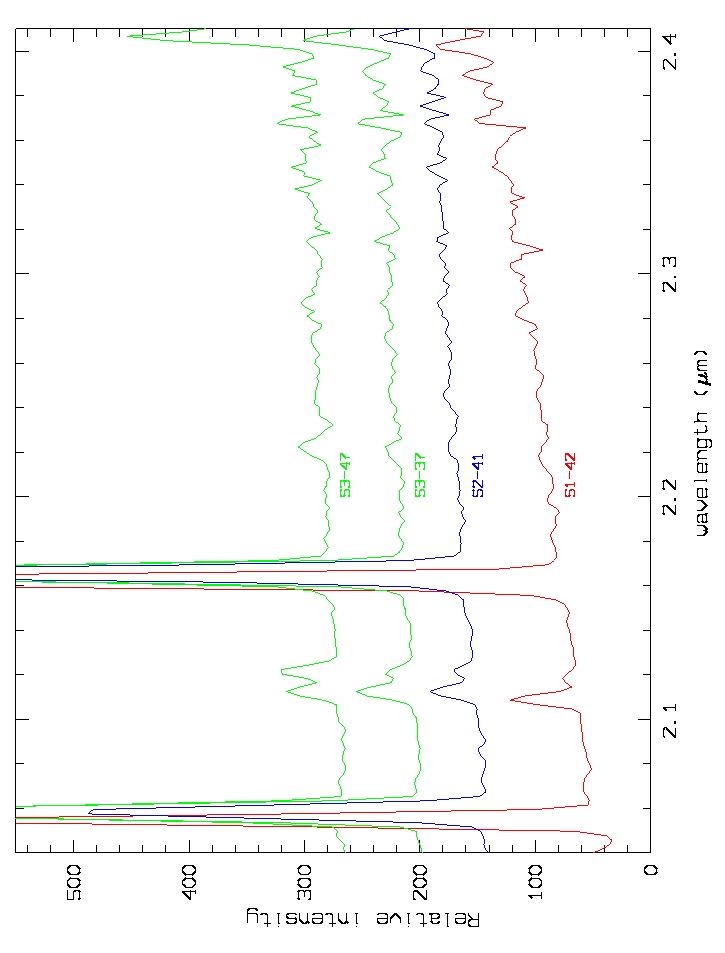}
\caption{One-dimensional  spectra of the stars  $\#$37, $\#$41,  $\#$42 and  $\#$47
         plotted in the range of 2.04-2.41 $\mu$m. }
\label{fig:spectresstar}
\end{center}
\end{figure}    
\begin{figure}[t]
\begin{center}
\hspace{0cm}
\vspace{0cm}
\includegraphics[width=8cm,height=9cm,angle=-90]{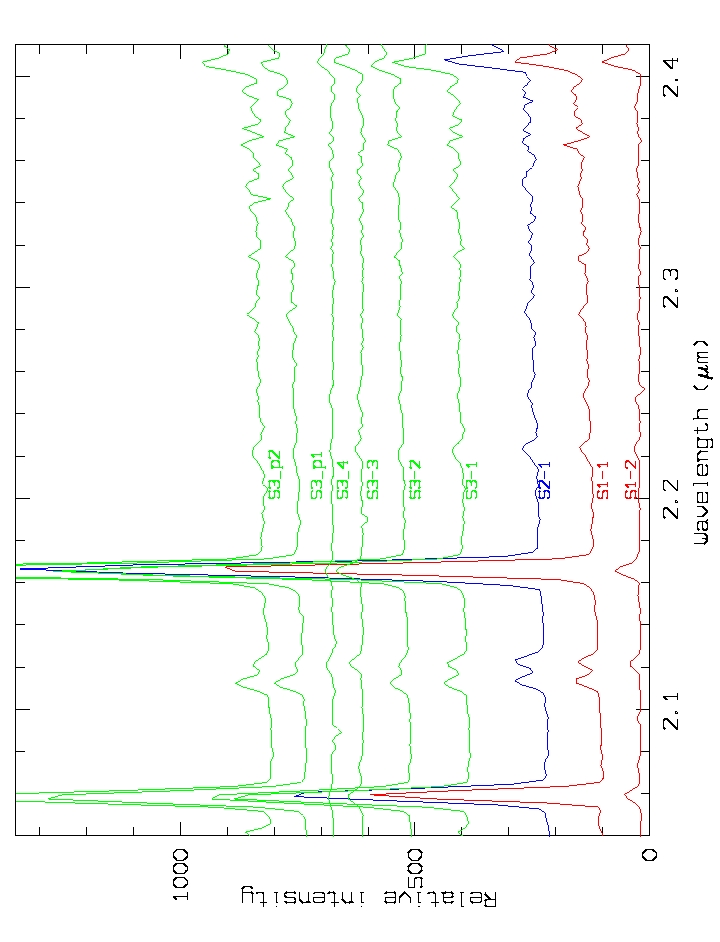}
\caption{One-dimensional  nebular spectra plotted in the range of 2.04-2.41 $\mu$m. }
\label{fig:spectresH2}
\end{center}
\end{figure}
\subsection{Search for YSO candidates in N88A and N88}
  Due to their IR-excess emission, YSOs are positioned in the redder parts of
 the CM and CC diagrams. We first examined low spatial
 resolution ($\sim$2.5$\arcsec$) mid-IR Spitzer data of N88A (Charmandaris et al. 2008),
 and then the near-IR stellar content of both the regions N88A and N88 obtained
 with the high spatial resolution allowed by NACO ($\sim$0.10 - 0.3$\arcsec$).  On the CM
 plot [3.6]-[8] versus [8], presented by Charmandaris et al. (2008), N88A lies at
 the border of the box representing the domain of Class II YSOs. N88A appears also 
inside the H II region domain. Similarly,  on the CC [5.8]-[8] versus [3.6]-[8]
 diagram N88A is located near the H II region domain, but outside Class I and
 Class II YSO areas. These observations are explained by the fact that N88A is above
 all a very bright H II region with strong nebular emission lines and affected by
 heavy extinction from local dust. In fact the Spitzer data represent flux integrations
 over the whole H II region ($\sim$1 pc$^{2}$).  Therefore, detecting a YSO inside the H II region
 seems hazardous unless the YSO is the dominant source inside N88A, which obviously is 
not the case. Note that although on the [5.8]-[8] versus [3.6]-[8] colour diagram, 
based on model calculations (Whitney et al. 2004), N88A appears among Class 0 and Class I
 data points. This diagram is not applicable to the case of N88A for the reasons explained earlier.   

In order to probe the presence of YSOs inside N88A we used our high resolution JHK data.
 The stars $\#$37, $\#$41, $\#$42 and $\#$47 in N88A-cl  are located at the upper part of the CM diagram
 (Fig. \ref{fig:HR}). In the JHK CC diagram (Fig. \ref{fig:HKJH}) these stars exhibit an H-K color ranging
 from 0.48 to 0.81 mag, and can be YSO candidates according to the JHK CC diagram
 of Maercker \& Burton (2005). However, their positions on the J-K versus K  diagram (Fig. \ref{fig:HR}) suggest
 heavily reddened main-sequence (MS) massive stars of masses between 15$M_{\sun}$ and 30 $M_{\sun}$. 
 Their positions between the 300 Myr and 1 Gyr isochrones are also compatible with supergiants. If confirmed
 as supergiants, these stars would not be physically associated
 with N88A, which is very young. The assumption of reddened MS massive stars or 
 massive YSOs seem more plausible. It is  very difficult to distinguish between these possibilities.
 Consequently, caution must be applied, using only
 JHK band observations to infer circumstellar material fraction in strong nebulous environments.
 The high spatial, but low spectral resolution S1, S2 and S3 spectra crossing the stars  $\#$37,
 $\#$41, $\#$42 and $\#$47 (see Fig. 10) do not allow to analyze accurately the Br$\gamma$〓 line emission profile,
 which is characteristic for YSO sources, and new spectroscopic observations are needed to clarify the
 nature of these stars. In N88A, L1 presents a relatively bright peak (L1-C) in the L'-band (Fig. 6).
 With K-L' $\ge$4.5, L1-C can be interpreted as a deeply embedded protostar (Lada et al. 2000). L1-C with
 a FWHM slightly larger than the PSF should still be in its contraction phase, surrounded by a dust
 shell. With a magnitude of 14 and a strong IR-excess (Table 4) L1-C could be a massive protostar of Class I.

 On what concerns the region outside N88A, from our high spatial resolution 
 data the JHK CC diagram (Fig. \ref{fig:HKJH}), taking into account the uncertainties excluding some stars
 close to the reddening vector, we show hereabove that in the extended N88 region, at least
 30\% of the faint detected stars have an IR-excess. These reddened stars seem to belong to a 
 cluster of faint stars labeled N88cl (Fig. 1) coinciding with the young cluster HW81
 (Hodge \& Wright 1977) formed of bright stars, situated towards N88 and not affected by dust
 (HM99). The JHK CC diagram (Fig. \ref{fig:HKJH}) shows that the bright stars $\#$6, $\#$8, $\#$11, $\#$12, $\#$15, $\#$28
 and $\#$59 in HW81 have no IR-excess and their masses spread in the range of 15M$\sun$ to 30M$\sun$.
 In N88-cl most of the reddened stars have a mass $\le$  12M$\sun$ (Fig. \ref{fig:HR})
 and are probably intermediate-mass YSO candidates. We assume that N88-cl belongs to N88
(Fig. 1). The stars $\#$21, $\#$33 and $\#$46 of N88-cl aligned on the east
filament of N88B (Fig. 1) with an H-K excess $>$ 0.7 mag, could be good
YSO candidates (Fig. \ref{fig:HKJH}). Their alignment suggests that their formation
may be triggered by the expansion of the shell around N88B.
\begin{figure}
\hspace{-0.3cm}
\includegraphics[width=13cm,height=15cm,angle=0]{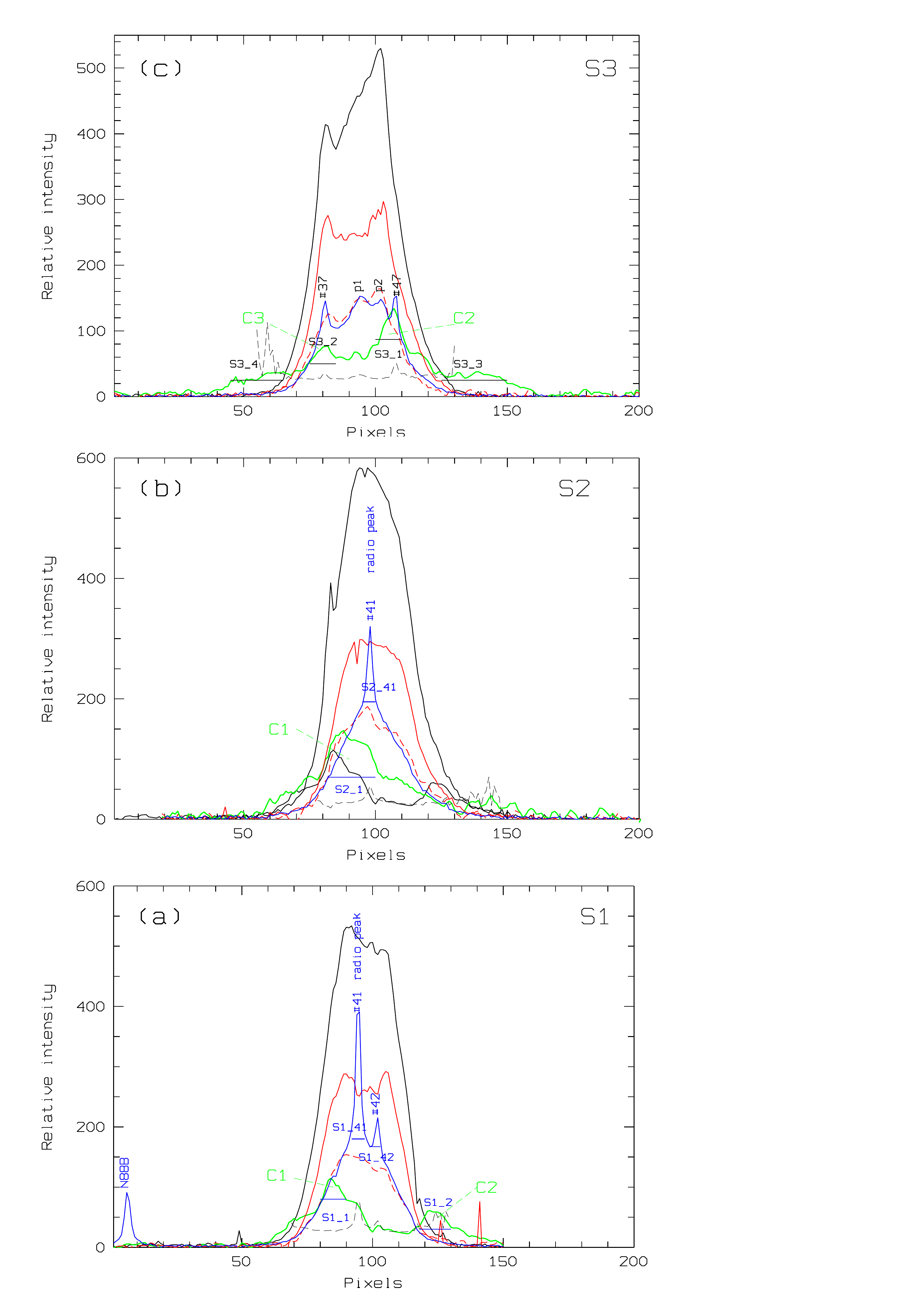}
\caption{ {\bf a)} Relative intensity distributions along slit S1 aligned on stars $\#$31,
          $\#$41 and $\#$42 (Fig. \ref{fig:S54inset}).
          {\bf b)} Relative intensity distributions along slit S2  crossing the star $\#$41. {\bf c)} 
          Relative intensity distributions along  slit S3 aligned on stars  $\#$37 and $\#$47.
          The red line represents the intensity distribution of the He I 2.058 $\mu$m
          emission line. The dashed red line represents the He I 2.113 $\mu$m  and the green one 
          the H$_{2}$ 2.121 $\mu$m, both are multiplied by a factor 5. The continuum/Br$\gamma$ ratio multiplied 
          by a factor 100 is plotted (dashed line). The position of each 1-D spectrum 
          is indicated by a small horizontal segment (solid line). 
          The plot range corresponds to 10.6 $\arcsec$  (1pix = 0.05273$\arcsec$). }
\label{fig:S1S2S3}
\end{figure} 

\subsection{The ionizing sources} 
 \subsubsection{N88A}
 At 3cm radio emission,  Indebetouw et al. (2004)   found for N88A a Lyman continuum flux of
 log N$_{Lyc}$ = 49.5. Using  the  spectral classification of  Smith et al. (2002) we estimate, 
 from this flux,  the  type of the ionizing source of N88A to range from O4 to O5.
 The type of the ionizing source derived by HM99 using the H$\beta$ flux corresponds to an O6 star.
  The  extracted 1-D spectra at  different positions  along the slits S1, S2 and S3 crossing
 N88A  (Fig. \ref{fig:S27inset}) are listed in Table 6. In each position the  rows are averaged 
  and the  corresponding 1-D spectra are shown in Fig. \ref{fig:spectresstar}
 for the stars  and  Fig. \ref{fig:spectresH2} for nebular emission. The positions of these
 spectra are represented by horizontal line segments on the plots corresponding to the
 distribution of the Br$\gamma$, He I 2.058, 2.113 and H$_{2}$ 2.121 $\mu$m emission lines as well
 as the continuum emission  (Figures \ref{fig:S1S2S3}a, b and c). The length of the segment is
 proportional to  the number of lines integrated along the slit. All the emission lines are
 continuum subtracted. This figure  also indicates the H$_{2}$ components  C1, C2 and C3
 (see Sect. 3.5) of which  the distribution  intends to clarify the presence of the structures
 seen in Figure 13 (see Sects. 3.6 and 3.7).  From the spectra presented in Figs. \ref{fig:spectresstar}
and Fig. \ref{fig:spectresH2}  we derive
  a ratio  He I 2.113$\mu$/Br$\gamma$ lines
 of mean value 0.06 indicating a hot O star of T$_{eff}$ $\geq$ 40000 K (Hanson et al. 2002).
 Table 6 (col. 8) shows that this ratio is fairly constant across N88A.

 In the spectra of stars  $\#$37, $\#$41, $\#$42 and $\#$47 (Fig. \ref{fig:spectresstar})
 the   He II 2.185$\mu$m  absorption is not detected, if present. The detection  is difficult  
 because of  our low signal/noise  and our low spectral resolution. The NIII 2.115$\mu$m is
 not detected either. 
 When the He II 2.185$\mu$m absorption line  is not present (Bik et al. 2005),
 the spectral type of a  star  should be later  than O8 V, which is the case for 
 our four resolved stars.
  As seen above,  the spectral type of the  ionizing source of the whole nebula derived from
 radio and H$\beta$ flux   ranges between O4 and O6. We  will adopt a type  O5 for our computation.
   Its comparison  with the type O8 V derived from our spectroscopy  for  the bright
 central star  $\#$2-41  clearly shows that other massive stars must contribute to
 the ionization of N88A. The flux  excess  between the ionizing star  $\#$41 of
 type O8 V and the O5 type derived from the flux  could be produced  by at least five O8 V stars.
 The massive stars $\#$37, $\#$42, and  $\#$47 (Fig. \ref{fig:S27zoom})
 located in the upper part of the CM diagram  could be good candidates for the ionization of N88A.
  The 3cm radio peak  centered at ($\alpha$, $\delta$) = (1$^h$24$^m$7$^s$.9, -73$^o$9$\arcmin$4$\arcsec$)
  and  our images show that the radio peak coincides perfectly  with  the central bright star
  $\#$41 (1$^h$24$^m$7$^s$.95, -73$^o$9$\arcmin$3$\arcsec$8) (Table 4). This strong radio
 emission superposing the NIR emission  Br$\gamma$ line   2.16$\mu$m line (Figs. \ref{fig:S1S2S3}a and b) 
 is  characteristic for an H II region. 

 \subsubsection{The N88A-cl cluster}
 On the JHK image of Lada et al. (2000) the Trapezium region of size of $\sim$0.75 x 0.75 pc
 located at a distance of  450 pc, contains four bright 
 central massive  stars and a plethora of low-mass stars with IR-excess. In our Ks-band 
 N88A has approximatively a similar diameter (Fig. \ref{fig:S27zoom}b) and  contains  also
 the  cluster N88A-cl. This cluster contains  other resolved stars not identified by DAOPHOT
 as well as  unresolved stars in crowded groups (Fig. \ref{fig:S27zoom}b). Among these
 stars the four brightest  ones analyzed using the JHK bands  also exhibit IR-excess. 
  N88A with its cluster  appears  morphologically comparable  with the Trapezium region, and
 other compact star formation regions of similar size, like SH2 269 in our galaxy, and N159-5
 in the LMC (Testor et al. 2007).
  N88A-cl can also be  compared with the pre-main-sequence (PMS) clusters, candidate YSOs,
 of size 0.24 pc to 2.4 pc found in SMC-N66 (Gouliermis et al. 2008).
 Like  N88A-cl these PMS clusters are found coinciding with
 [OIII], H$\alpha$ and H$_{2}$ emission peaks (see Sect. 3.7).
 Their clustering properties are  similar to the star forming  region  Orion despite its
 higher  metallicity (Hennekemper et al. 2008).
 \begin{figure*}[t]
\begin{center}
\hspace{0cm}
\includegraphics[width=15cm]{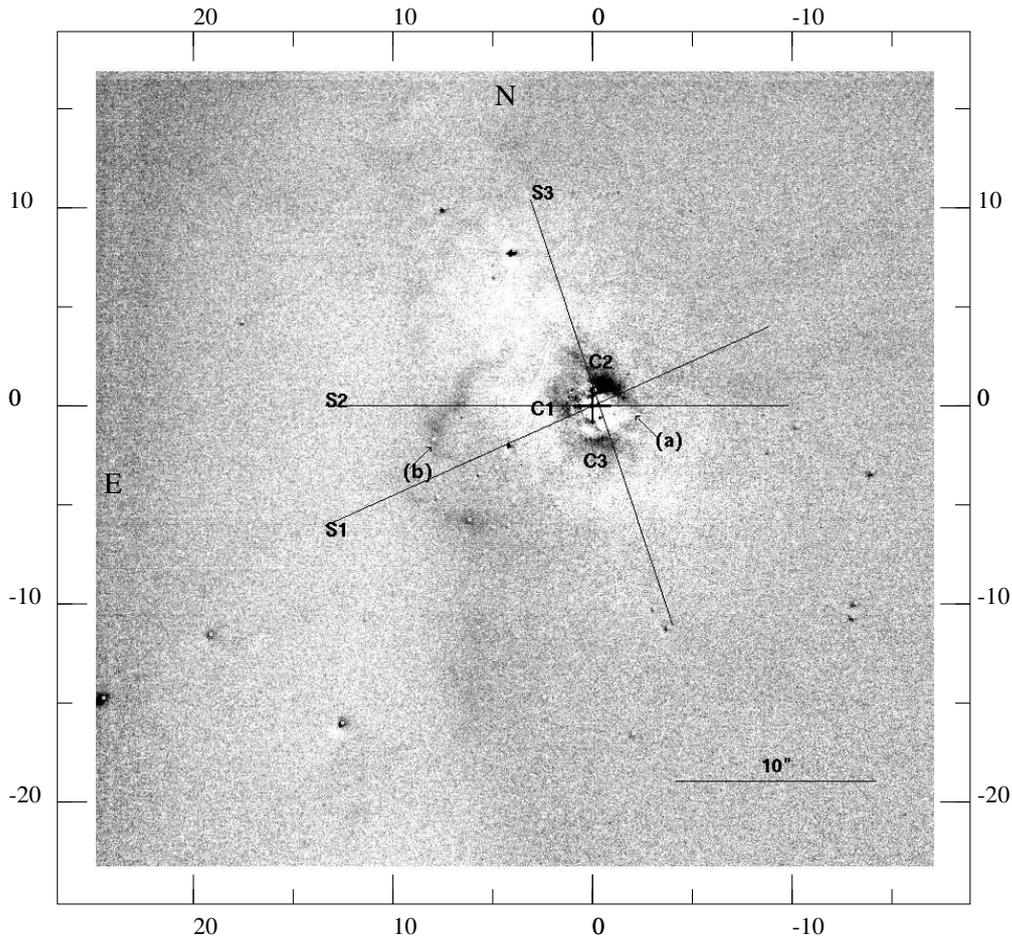}
\vspace{-6.9cm}
\caption{The H$_{2}$ 2.121 $\mu$m emission (continuum subtracted)  showing the three components
        C1, C2 and C3 of the shell (a) as well as the larger shell (b).
 The R.A. and Dec. offsets (arcsec) are from star $\#$41 represented by a cross (coordinates 0,0).
 The image size  is 42$\arcsec$ x 40$\arcsec$, corresponding to $\sim$ 13 x 12 pc. }
\label{fig:dif}
\end{center}
\end{figure*}  
  \subsubsection{The bright central star  $\#$41}

 In the Trapezium, the four ionizing bright stars lie within a diameter of $\sim$ 0.05 pc. At
 the  distance of the SMC our spatial resolution is $\sim$ 0.03 pc ($\sim$ 6900 AU).
 With this relative lower spatial resolution it is not excluded that the ionizing
 star  $\#$41 could also be a tight young cluster. This assumption is strengthened by
 the photometry. The magnitude of this star obtained using the PSF  is 15.05 (Table 4).
 Its dereddened magnitude derived with
 A$_{K}$ = 0.58 mag corresponds to a mass of $\sim$ 40 $M_{\sun}$ (Fig. \ref{fig:HR}). 
 Using the parameters for O stars of Vacca (1996), we  classify  $\#$41 as a O6.5 V type star
 instead of O8 V derived by spectrocopy, and it could also be multiple.

\renewcommand{\arraystretch}{1}
\tabcolsep=0.06cm
 \begin{table*}
\caption{ Relative integrated  HeI, HI and   H$_2$ emission-line fluxes  in N88A for  14 positions 
 extracted across S1, S2 and S3  (Fig. \ref{fig:S1S2S3}abc). }
\begin{center}
\begin{tabular}{l l l l l l l c }
\hline
\noalign{\smallskip}
Id     &  HeI    &  HeI     & H$_{2}$ 1-0 S(1) &Br$\gamma$ &H$_{2}$ 1-0 S(0) &H$_{2}$ 2-1 S(1) &\hspace{0.2cm} I(He 2.113$\mu$m)/I(Br$\gamma$)\\
       &2.058$\mu$m&2.113$\mu$m &2.121$\mu$m  &2.166$\mu$m  &2.223$\mu$m &2.247$\mu$m & \\
\hline
\noalign{\smallskip}
S1-41  & 3.18        & 0.375         &  0.121    &  6.07        &  0.099   &  0.031   &0.062\\
S1-42  & 3.36        & 0.337         &  0.075    &  5.94        &  0.092   &  0.044   &0.057\\
S2-41  & 2.38        & 0.257         &  0.094    &  4.76        &  0.068   &  0.038   &0.054\\
S3-37  & 0.76        & 0.065         &  0.028    &  1.21        &  0.018   &  0.009   &0.054\\
S3-47  & 0.60        & 0.055         &  0.066    &  1.01        &  0.033   &  0.006   &0.054\\
S1-1   & 2.43        & 0.24          &  0.199    &  4.30        &  0.126   &  0.069   &0.054\\
S1-2   & 0.17        & 0.017         &  0.094    &  0.31        &  0.054   &  0.035   &0.057\\
S2-1   & 1.49        & 0.16          &  0.150    &  3.24        &  0.093   &  0.048   &0.049\\
S3-1   & 0.67        & 0.065         &  0.049    &  1.11        &  0.033   &  0.009   &0.059\\
S3-2   & 0.58        & 0.057         &  0.026    &  0.95        &  0.018   &  0.009   &0.060\\
S3-3   & 0.02        & 0.002         &  0.017    &  0.04        &  0.010   &  0.008   &0.056\\
S3-4   & 0.02        &               &  0.019    &  0.02        &  0.009   &  0.009   &\\
S3-p1  & 0.74        & 0.092         &  0.020    &  1.42        &  0.014   &  0.009   &0.065\\
S3-p2  & 0.89        & 0.090         &  0.036    &  1.55        &  0.034   &  0.008   &0.058\\
\hline
\noalign{\smallskip}
\end{tabular}
\end{center}
\end{table*} 

\subsection{The continuum dust emission}
 Through  the JHK-band, N88A presents a relatively bright nebular continuum emission centered on
 star $\#$41 (Figs. \ref{fig:S27zoom}a and \ref{fig:PSF}). In the L'-band the continuum is less 
 homogeneous, due to the brightness of L1, L2, L3  and L4.  Although the signal to noise ratio is not high,
 L4 appears very faint in the y-band image (Fig. \ref{fig:S27zoom}c).
 The nature of this continuum  is not clear. The relatively
 strong L-band excess of K-L' = 2 mag derived from the integration of the whole region
 N88A (Table 4) supposes that the  continuum  could come  from  the emission
 of  circumstellar material (Lada et al. 2000) around resolved and unresolved young
 stars. These stars could be located mainly at the positions L1, L2, L3 and L4 (Fig. \ref{fig:S27zoom}d).
 L4 shows two peaks P1 and P2 visible on the continuum plot of S3 (Fig. \ref{fig:S1S2S3}c).
 However, the continuum  emission could also be formed by  interstellar dust associated
 with the gas of the CO cloud (Testor et al. 1985, HM99, Stanamirovic et al. 2000, Israel et al. 2003).
 In Fig. \ref{fig:S27zoom}c the strong optical emission  at the  position of the `dark spot'
 could be explained by strong dust scattering reflecting the light of at least  star  $\#$37.
  The  nature of the continuum emission of N88A should be   a combination of the
 two possibilities: emission of  circumstellar material and/or  dust associated
 with the gas. Along the  slits
 S1, S2 and S3 the intensity distribution of the continuum near the Br$\gamma$ line shows
 a strong continuum/Br$\gamma$  ratio of 0.30-0.4 over a range of 6$\arcsec$ around star
  $\#$41  (Figs. \ref{fig:S1S2S3}ab).  In these figures the broadness of the continuum
 and Br$\gamma$ distribution  are similar (FWHM $\sim$ 1.8$\arcsec$). 

\subsection{H$_2$ emission}

 In Testor et al. (2005) the  profile of the H$_2$ emission along the slit  corresponding to 
 our spectrum  S1 appears in the form of two  blended profiles, due to the  low spatial
 resolution spectroscopy.
 Thanks to the high  spatial resolution of our new data, the  complex morphology of the 
 H$_2$ emission in N88A is revealed both  by imagery and spectroscopy. 
 Fig. \ref{fig:dif} shows a bidimensional image of the H$_2$ emission (v=1-0 S(1) line).
 This image is achieved by subtracting the image in the 2.24 $\mu$m filter, which allows
 the passage only of the continuum radiation from the image in the 2.13 $\mu$m filter. 
  In this H$_2$ image, N88A resembles a circular shell (a) of diameter $\sim$ 3$\arcsec$
 with three maxima labelled C1, C2 and C3 (Fig. \ref{fig:dif}). Within (a) there is  
 a cavity  suggesting radiation pressure, especially from the four central stars.
 The structure C2 is very bright and extended along the direction northeast-southwest
 and coincides with the ionization front detected by HM99. C2 has a  sharp extension in
 the direction of star  $\#$47. The  H$_2$-band image also shows that the bow-shaped
 filament located east of N88A  seems to belong to a second  shell (b)
 of  diameter 7$\arcsec$ centered on N88B (Fig. \ref{fig:dif}). The shells (a) and (b) seem 
 to be in interaction approximatively at the position C1.

  Unlike the spectra obtained with ISAAC (Testor et al. 2005) the high-spatial resolution
 long-slit spectra S1, S2 and S3 allow to resolve the  inner structures  and stars of  N88A.
 In the direction of the slit S1, the two H$_2$ emission structures C1 and C2 
 are well resolved (Fig. \ref{fig:S1S2S3}a) and separated by $\sim$ 2$\arcsec$. 
 C1 distant of $\sim$ 0.5 $\arcsec$ from  $\#$41  coincides  with the absorption
 lane observed in optical images by Kurt et al. (1999) and HM99. In the direction of the 
 slit S2, only the structure C1 is seen east of star $\#$41 (Fig. \ref{fig:S1S2S3}b). 
 In the direction of the slit  S3, the well seen structures C2 and C3 (Fig. \ref{fig:S1S2S3}c)
 coincide with the stars  $\#$47 and  $\#$37 respectively. 
  According to Rubio et al. (2000) massive star formation could be taking place in
  dense H$_{2}$ knots associated with molecular clumps. According to Gouliermis et al. (2008)
  PMS clusters could be  candidate YSOs. These results  strengthen  our assumption that
 N88A-cl could contain YSOs.

\begin{figure*}
\begin{center}
\hspace{0.cm}
\includegraphics[width=14cm,height=10cm,angle=0]{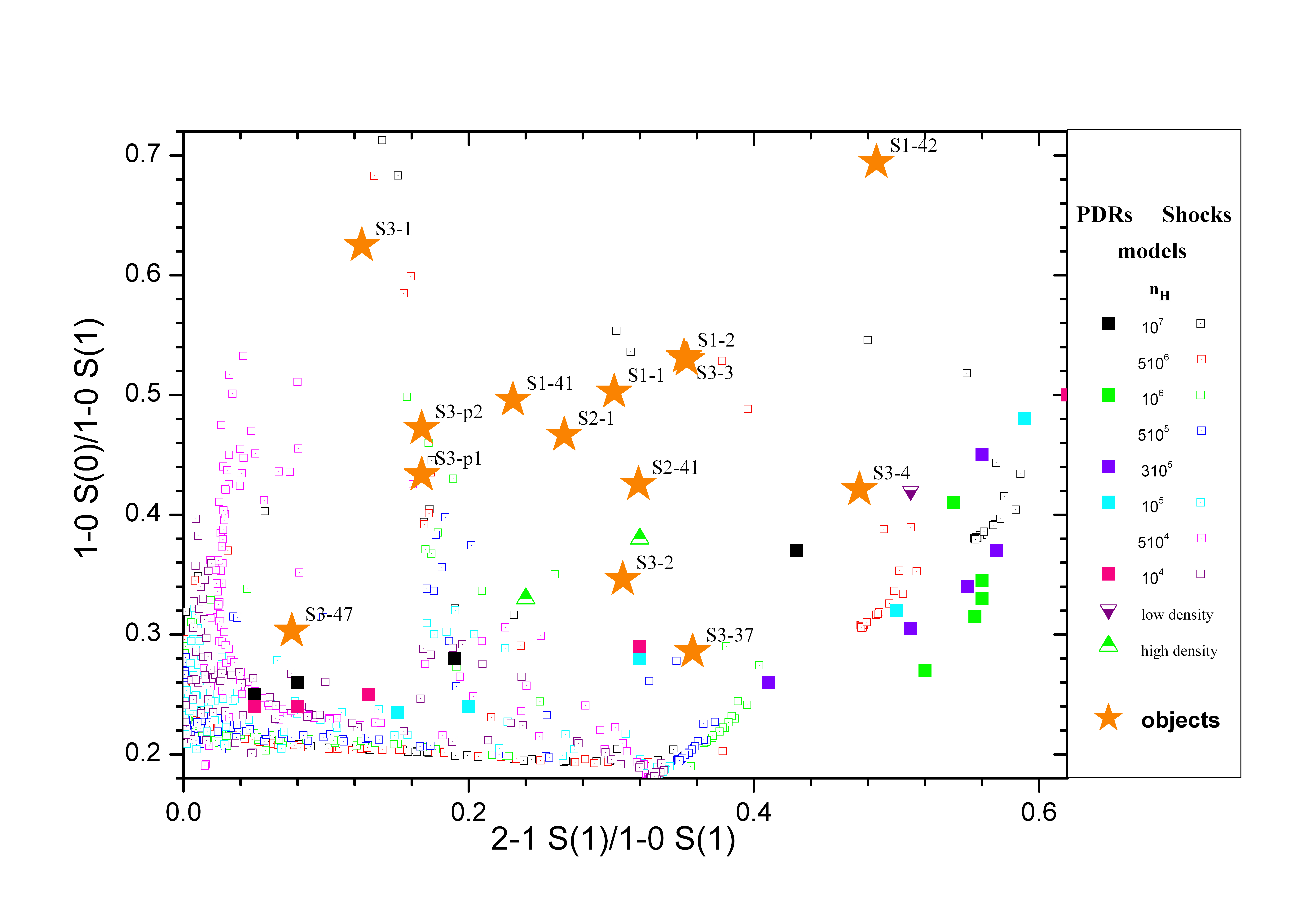}
\caption{2-1 S(1)/1-0 (S1) vs. 1-0 S(0)/1-0 S(1) ratio at different positions in the HEB N88A.}
\label{fig:tableline}
\end{center}
\end{figure*}

 From Israel $\&$ Koorneef (1988), the molecular hydrogen emission may be caused either by shock
 excitation due to embedded stars,  or by fluorescence  of molecular material in the
 ultraviolet radiation field  of the OB stars exciting the H II region in the molecular cloud.
 They  conclude that in the MCs, shock excitation of H$_{2}$ is only expected very close 
 to (i.e. 0.15pc) stars embedded in a molecular cloud. At a larger distance,
 radiative excitation of H$_{2}$ by the UV radiation field  of the OB stars is the only mechanism.  
 Their spectrophotometry with a large aperture (10$\arcsec$)  made difficult a precise
 determination of the brightness of the lines   2-1(S1), 1-0(S1) and 1-0(S0) usually considered
 between  shocks and radiative excitation. These lines deblended when necessary  have been
 derived from our low-spectral-resolution spectra S1, S2 and S3 crossing the  zones C1, C2 and C3
 as well as the stars
  $\#$41, $\#$42, $\#$37 and $\#$47 and their intensities are shown in Table 6.
 None of these lines suffers from atmospheric absorption, considering a V$_{lsr}$ of
 147 km s$^{-1}$  (Israel et al. 2003), as it can be  derived from the solar spectrum
 atlas (Livingston \& Wallace 1991) with the help of a useful home-made software$^{1}$\footnotetext[1]
 {http://www.u-cergy.fr/LERMA-LAMAP/informatique/raiesH2/index.html}.
 The lines may suffer from differential reddening. Mathis (1990) estimates  that the effect for
 Galactic Sources follows a power-law in the J-, H-, and K-bands:
 $I_{1}/I_{2} = (\lambda_{1}/\lambda_{2})^{-1.7}$.
 The effect on the v=1-0 S(0) and  v=2-1 S(1)  lines would be that they are overestimated  by 10\%.
 However, this is based entirely on Galactic Sources. To the best of our knowledge, a differential
 reddening law has not been determined for Extra-Galactic Sources. Moreover, since the effect is already 
 within our observational uncertainty, we choose to ignore it.
 In Figure \ref{fig:tableline} the strengths of the  2.247$\mu$m 2-1(S1)  and 2.223$\mu$m 1-0(S1)
 lines are shown relative to the 2.121$\mu$m 1-0(S1) line for all the objects in Table 6. For
 radiative excitation (PDRs), the usual criteria are that these ratios should range from 0.5 to 0.6
 and from 0.4 to 0.7 respectively (Black $\&$ van Dishoeck 1987), while for shock-excitation with
 T=2000K they should be 0.08 and 0.21 respectively (Shull $\&$ Hollenbach 1978). We  show in
 Figure \ref{fig:tableline} the results from Draine \& Bertoldi (1996) reported by Hanson et al.
 (2002) for high density (n$_H$=10$^6$ and $\chi$=10$^4$ and 10$^5$) and low density (n$_H$=10$^4$ and 
 $\chi$=10$^2$) PDRs. We also show in this figure results for more recent and elaborated 
 PDR models (Le Petit et al. 2006) [for n$_H$ ranging from 10$^4$ to 10$^7$ and $\chi$ 
 from 10$^3$ to 10$^7$] as well as for shock models (Flower $\&$ Pineau des For\^{e}ts 2003) [for 
 the same range of n$_H$,  v$_S$ from 10 to 60 km s$^{-1}$, the magnetic scaling factor b 
 from 0 to 10 and an ortho/para ratio of 3]. None of these models, either PDRs or shocks really 
 fit with our observations, with the exception of the objects S3-p1 and p2 as well as S3-1, well 
 inside the nebulosity, which could fit with shocks (relatively high velocity v$_S$, 30 to 50 km 
 s$^{-1}$) and low b $\sim$ 0.1. Nevertheless, apart from the three H$_{2}$ lines mentioned 
 above, four additional ones are observed: 2-1 S(2), 2-1 S(3), 3-2 S(1) and especially 3-2 S(2) at 
 respectively 2.154, 2.073, 2.386 and 2.286$\mu$m. These lines are more sensitive to 
 absorption by atmospheric lines, depending in fact on the accuracy of the v$_{lsr}$. The first 
 one is the less affected by positive or negative velocity shift, whilst the two following 
 are slightly absorbed up to 180 km s$^{-1}$ but may suffer a 50 $\%$ absorption at 140 km 
 s$^{-1}$. The last one is free of absorption between 115 and 150 km s$^{-1}$ and does not 
 appear to be blended with  any other lines. Possible turbulence in the emitting region may 
 broaden the lines, then lowering the effect due to atmospheric absorption. In any case the 
 concomitant appearance of lines emanating from high v or J as the 3-2 S(2) H$_2$ line shows 
 a clear trend in favor of the major presence of PDR excitation for most of the observed 
 objects, without nevertheless totally excluding the additional presence of shock excitation. 
 Clearly higher, both spatial and spectral resolutions are required to progress in the knowledge 
 of these objects.
\section{Conclusions}
 We present high spatial resolution imaging of  FWHM $\sim$ 0.12$\arcsec$ - 0.25$\arcsec$
 in the JHKL'-band  of  the HEB N88A and its immediate environment, and the main results
 are as follows: 

 N88A is associated with a cluster  that contains at least  thirteen stars  
 centered approximatively on the bright central star  $\#$41, that could be multiple.

 N88A coincides  perfectly with the 3cm radio peak and should be ionized not only by the 
 star   $\#$41  classified  of  type O8 V,  but also by   other low to high-mass stars.
 
 From analysis of the JHK CC diagram  we found four possible MYSO star candidates 
 in  the  N88A cluster, as well as three probable YSOs in the red bow east of N88A.
 In N88 at least 30$\%$ of the  detected stars have an IR-excess.

 From  the K-L excess we  found that the core L1-C in N88A  should be
 a heavily embedded, high mass protostar of Class I.
  
 The continuum emission at the position of  $\#$41 is very bright and represents  about 30$\%$
 of the  Br$\gamma$ emission peak.

 The H$_2$ infrared emission in N88A resembles a shell formed mainly by three peaks of which
 one coincides with the ionization front.

 We show that the excitation mechanism may be  caused predominantly by PDRs,
 without excluding combination with shocks.

 The morphology   of N88A could be comparable with  galactic   regions  such as 
 the nearby Trapezium region in the Orion nebula.   

 Future JHK band  imaging data, using   higher spatial resolution  and  longer wavelengths,
 as L' and M'  provided by the NACO S13 camera  are still needed to  disentangle the IR-excess origin
 in N88A. Higher spectral  resolution spectra are also required to obtain a better analysis of the
 different spectral lines like the Br$\gamma$  emission.  These new observations  should  allow to  investigate
 more  thoroughly  the HEB N88A. This object, which is the brightest, the most excited and reddened of the 
 MCs,  presents a unique opportunity  to progress in the knowledge of newborn  massive stars
 in regions of low metallicity. 

\begin{acknowledgements}
 We would  like to thank the Directors and Staff of the ESO-VLT for making possible these
 observations and particularly the NACO team for their excellent support. JLL, LK and SD would like 
 to acknowledge  the support of~the French PCMI program "Physico Chimie du Milieu Interstellaire",
 funded by the CNRS. This research has made use of the Simbad database, VizieR and
 Aladin operated at CDS, Strasbourg, France, and the NASA's Astrophysics Data System Abstract Service.
\end{acknowledgements}

\end{document}